\documentclass{emulateapj}
\usepackage{graphicx}
\usepackage{natbib}
\begin{document}
\newcommand{\lya}{Ly$\alpha$ }
\newcommand{\z}{\emph{z} $\sim$ }
\newcommand{\ewlya}{\emph{W}$_{\rm Ly\alpha}$}
\newcommand{\kms}{km s$^{-1}$}
\newcommand{\asec}{\mbox{$''\mskip-7.6mu.\,$}}
\newcommand{\amin}{\mbox{$'\mskip-7.6mu.\,$}}
\makeatletter
\newcommand{\Rmnum}[1]{\expandafter\@slowromancap\romannumeral #1@}
\makeatother

\title{THE RELATIONSHIP BETWEEN STELLAR POPULATIONS AND \lya EMISSION IN LYMAN BREAK GALAXIES\altaffilmark{1}}
\author{\sc Katherine A. Kornei and Alice E. Shapley\altaffilmark{2,}\altaffilmark{3}}
\affil{Department of Physics and Astronomy, 430 Portola Plaza, University of California at Los Angeles, Los Angeles, CA 90025, USA}
\author{\sc Dawn K. Erb\altaffilmark{4}}
\affil{Department of Physics, University of California at Santa Barbara, Santa Barbara, CA 93106, USA}
\author{\sc Charles C. Steidel}
\affil{California Institute of Technology, MS 105-24, Pasadena, CA 91125, USA}
\author{\sc Naveen A. Reddy\altaffilmark{5}}
\affil{National Optical Astronomy Observatory, 950 N. Cherry Ave, Tucson, AZ 85719, USA}
\author{\sc Max Pettini}
\affil{Institute of Astronomy, Madingley Road, Cambridge, CB3 0HA, UK}
\author{\sc Milan Bogosavljevi\'c}
\affil{California Institute of Technology, MS 105-24, Pasadena, CA 91125, USA}

\altaffiltext{1}{Based, in part, on data obtained at the 
W.M. Keck Observatory, which is operated as a scientific partnership among the California Institute of Technology, the University of California, and NASA, and was made possible by the generous financial
support of the W.M. Keck Foundation.} 
\altaffiltext{2}{Packard Fellow.}
\altaffiltext{3}{Alfred P. Sloan Fellow.}
\altaffiltext{4}{Spitzer Fellow.}
\altaffiltext{5}{Hubble Fellow.}

\begin{abstract}
We present the results of a photometric and spectroscopic survey of 321 Lyman break galaxies (LBGs) at \z3 to investigate systematically the relationship between \lya emission and stellar populations. \lya equivalent widths ($W_{\rm Ly \alpha}$) were calculated from rest-frame UV spectroscopy and optical/near-infrared/\emph{Spitzer} photometry was used in population synthesis modeling to derive the key properties of age, dust extinction, star formation rate (SFR), and stellar mass. We directly compare the stellar populations of LBGs with and without strong \lya emission, where we designate the former group ($W_{\rm Ly \alpha}$ $\ge$ 20 \AA) as Ly$\alpha$-emitters (LAEs) and the latter group ($W_{\rm Ly \alpha}$ $<$ 20 \AA) as non-LAEs. This controlled method of comparing objects from the same UV luminosity distribution represents an improvement over previous studies in which the stellar populations of LBGs and narrowband-selected LAEs were contrasted, where the latter were often intrinsically fainter in broadband filters by an order of magnitude simply due to different selection criteria. Using a variety of statistical tests, we find that \lya equivalent width and age, SFR, and dust extinction, respectively, are significantly correlated in the sense that objects with strong \lya emission also tend to be older, lower in star formation rate, and less dusty than objects with weak \lya emission, or the line in absorption. We accordingly conclude that, within the LBG sample, objects with strong \lya emission represent a later stage of galaxy evolution in which supernovae-induced outflows have reduced the dust covering fraction. We also examined the hypothesis that the attenuation of \lya photons is lower than that of the continuum, as proposed by some, but found no evidence to support this picture. 
\end{abstract}


\section{Introduction}

An increasing number of high-redshift galaxies have been found in the last two decades using selection techniques reliant on either color cuts around the Lyman limit at 912 \AA\ in the rest frame \citep[e.g.,][]{steidel1996a,steidel1999} or strong \lya line emission \citep[e.g.,][]{cowie1998,rhoads2000,gawiser2006}. These two methods, which preferentially select Lyman break galaxies (LBGs) and Ly$\alpha$-emitters (LAEs), respectively, have successfully isolated galaxies at redshifts up to $z$ = 7 \citep{iye2006,bouwens2008}. Extensive data sets of LBGs and LAEs have afforded detailed studies of galactic clustering \citep[e.g.,][]{adelberger1998,giavalisco2001}, the universal star formation history \citep[e.g.,][]{madau1996,steidel1999}, and the galaxy luminosity function \citep[e.g.,][]{reddy2008,mclure2009}. While the nature of LBGs and LAEs have been studied at a range of redshifts \citep[e.g.,][]{shapley2001,gawiser2006,verma2007,pentericci2007,nilsson2009,ouchi2008,finkelstein2009}, the epoch around \z3 is particularly well-suited to investigation of these objects' detailed physical properties. At this redshift, the prominent H\Rmnum{1} \lya line ($\lambda_{\rm rest}$ = 1216 \AA), present in all LAE spectra and a significant fraction of LBG spectra, is shifted into the optical, where current imaging and spectroscopic instrumentation is optimized. Consequently, there are large existing data sets of spectroscopically-confirmed \z3 LBGs \citep[e.g.,][]{steidel2003} and LAEs \citep[e.g.,][]{lai2008}, where extensive multiwavelength surveys often complement the former and, less frequently, the latter.

The mechanism responsible for LAEs' large \lya equivalent widths is not fully understood, although several physical pictures have been proposed \citep[e.g.,][]{dayal2009,kobayashi2010}. As \lya emission is easily quenched by dust, one explanation for LAEs is that they are young, chemically pristine galaxies experiencing their initial bursts of star formation \citep[e.g.,][]{hu1996,nilsson2007}. Conversely, LAEs have also been proposed to be older, more evolved galaxies with interstellar media in which dust is segregated to lie in clumps of neutral hydrogen  surrounded by a tenuous, ionized dust-free medium \citep{neufeld1991,hansen2006,finkelstein2009}. In this picture, \lya photons are resonantly scattered near the surface of these dusty clouds and rarely encounter dust grains. Continuum photons, on the other hand, readily penetrate through the dusty clouds and are accordingly scattered or absorbed. This scenario preferentially attenuates continuum photons and enables resonant \lya photons to escape relatively unimpeded, producing a larger \lya equivalent width than expected given the underlying stellar population. To date, the distribution of dust in the interstellar medium has only been investigated using relatively small samples \citep[e.g.,][]{verhamme2008,atek2009,finkelstein2009}.

Given the different selection techniques used to isolate LBGs and LAEs, understanding the relationship between the stellar populations of these objects has been an important goal of extragalactic research. Recent work by \citet{gawiser2006} has suggested that LAEs are less massive and less dusty than LBGs, prompting these authors to propose that LAEs may represent the beginning of an evolutionary sequence in which galaxies increase in mass and dust content through successive mergers and star formation episodes \citep{gawiser2007}. The high specific star formation rate -- defined as star formation rate (SFR) per unit mass -- of LAEs \citep[$\sim$ 7 $\times$ 10$^{-9}$ yr$^{-1}$;][]{lai2008} relative to LBGs \citep[$\sim$ 3 $\times$ 10$^{-9}$ yr$^{-1}$;][]{shapley2001} illustrates that LAEs are building up stellar mass at a rate exceeding that of continuum-selected galaxies at \z3. This rapid growth in mass is consistent with the idea that LAEs represent the beginning of an evolutionary sequence of galaxy formation. However, results from \citet{finkelstein2009} cast doubt on this simple picture of LAEs as primordial objects, given that these authors find a range of dust extinctions ($A_{\rm 1200}$ = 0.30--4.50) in a sample of 14 LAEs at \z4.5. \citet{nilsson2009} also find that \z2.25 LAEs occupy a wide swath of color space, additional evidence that not all LAEs are young, dust-free objects. Furthermore, the assertion that LAEs are pristine galaxies undergoing their first burst of star formation is called into question by the results of \citet{lai2008}. These authors present a sample of 70 \z3.1 LAEs, $\sim$ 30\% of which are detected in the 3.6 $\mu$m band of the \emph{Spitzer} Infrared Array Camera \citep[IRAC;][]{fazio2004}. These IRAC-detected LAEs are significantly older and more massive ($\langle t_{\star} \rangle$ $\sim$ 1.6 Gyr, $\langle M \rangle$ $\sim$ 9 $\times$ 10$^{9}$) than the IRAC-undetected sample ($\langle t_{\star} \rangle$ $\sim$ 200 Myr, $\langle M \rangle$ $\sim$ 3 $\times$ 10$^8$ $M_{\odot}$); \citet{lai2008} suggest that the IRAC-detected LAEs may therefore be a lower-mass extension of the LBG population. Narrowband-selected LAEs are clearly marked by heterogeneity, and the relationship between these objects and LBGs continues to motivate new studies. 

When comparing the stellar populations of LBGs and LAEs, it is important to take into account the selection biases that result from isolating these objects with broadband color cuts and line flux/equivalent width requirements, respectively. By virtue of selection techniques that rely on broadband fluxes and colors, LBGs generally have brighter continua than LAEs. Spectroscopic samples of LBGs typically have an apparent magnitude limit of ${\cal R}$ $\le$ 25.5 \citep[0.4$L^*$ at \z3;][]{steidel2003} while LAEs have a median apparent magnitude of $R$ $\sim$ 27 \citep{gawiser2006}, where ${\cal R}$ and $R$ magnitudes are comparable. Even though the majority of LBGs studied to date are an order of magnitude more luminous in the continuum than typical LAEs, both populations have similar rest-frame UV colors \citep{gronwall2007}. Therefore, LAEs fainter than ${\cal R}$ = 25.5 are excluded from LBG spectroscopic surveys not because of their colors, but rather because of their continuum faintness. Given the significant discrepancy in absolute magnitude between LBGs and LAEs, understanding the relationship between these objects can be fraught with bias. An preferable approach to comparing these populations is to investigate how the strength of \lya emission is correlated with galaxy parameters, for a controlled sample of objects at similar redshifts \emph{drawn from the same parent UV luminosity distribution}.

Several authors have looked at the question of the origin of \lya emission in UV flux-limited samples \citep[e.g.,][]{shapley2001,erb2006a,reddy2008,pentericci2007,verma2007}. \citet{shapley2001} analyzed 74 LBGs at \z3 and constructed rest-frame UV composite spectra from two samples of ``young" ($t_{\star}$ $\leq$ 35 Myr) and ``old" ($t_{\star}$ $\geq$ 1 Gyr) galaxies, respectively.  These authors found that younger objects exhibited weaker \lya emission than older galaxies; \citet{shapley2001} attributed the difference in emission strength to younger LBGs being significantly dustier than their more evolved counterparts. On the other hand, \citet{erb2006a} examined a sample of 87 star-forming galaxies at \z2 and found that objects with lower stellar mass had stronger \lya emission features, on average, than more massive objects. In a sample of 139 UV-selected galaxies at \z2--3, \citet{reddy2008} isolated 14 objects with \lya equivalent widths $\ge$ 20 \AA\ and noted no significant difference in the stellar populations of strong Ly$\alpha$-emitters relative to the rest of the sample. \citet{pentericci2007} examined 47 LBGs at \z4 and found that younger galaxies generally showed \lya in emission while \lya in absorption was associated with older galaxies (in contrast to the \citet{shapley2003} results). Probing even earlier epochs, \citet{verma2007} examined a sample of 21 LBGs at \z5 and found no correlation between \lya equivalent width and age, stellar mass, or SFR. These authors noted, however, that only 6/21 of the brightest LBGs had corresponding spectroscopy from which equivalent widths were estimated. Therefore, the lack of a correlation between \lya equivalent width and stellar populations may, in this case, have been masked by a small sample that was biased towards the brightest objects. These aforementioned investigations have shown that there does not yet exist a clear picture relating stellar populations to \lya emission. 

\begin{figure}[t]
\centering
\includegraphics[trim=0in 0in 0in 0in,clip,width=3in]{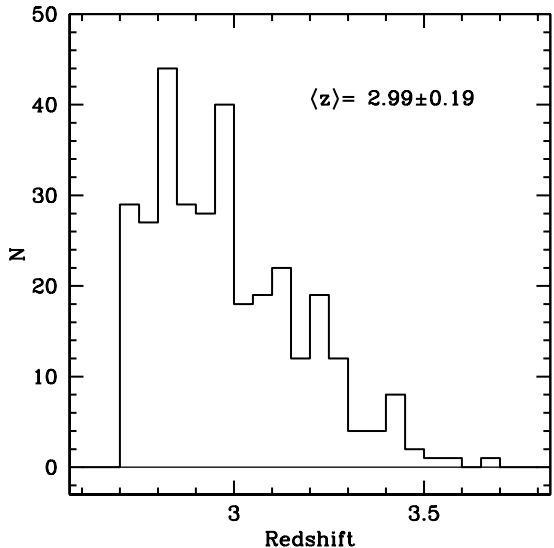}
\caption{Redshift distribution of the sample, where $\langle z \rangle$ = 2.99 $\pm$ 0.19.}
\label{fig: redshift} 
\end{figure}

In this paper, we present a precise, systematic investigation of the relationship between \lya emission and stellar populations using our large photometric and spectroscopic data set of \z3 observations. As an improvement over previous studies, we approach the data analysis from multiple aspects: we compare not only the stellar population parameters derived from population synthesis modeling, but also examine the objects' best-fit SEDs and photometry. Furthermore, all analysis is conducted on objects drawn from the same parent sample of continuum-bright (${\cal R}$ $\le$ 25.5) LBGs. By controlling for continuum magnitude, we avoid the biases of comparing objects with significantly different luminosities while still retaining the ability to comment on the nature of strong Ly$\alpha$-emitting galaxies within the LBG sample. Our conclusions are applicable to both LBGs and bright ({$\cal R$} $\le$ 25.5) narrowband-selected LAEs. While we are unable to make inferences about the population of faint LAEs, our study is complete with respect to bright LAEs given these objects' similar colors to LBGs in the rest-frame UV. 

We are motivated by the following questions: how do the stellar populations of Ly$\alpha$-emitting LBGs differ from those of other LBGs at \z3 where the \lya emission line is weaker (or absent altogether)? To what degree are galactic parameters such as dust extinction, SFR, age, and stellar mass correlated  with \lya line strength? What do the relative escape fractions of \lya and continuum photons reveal about the distributions of gas and dust in these objects' interstellar media?

This paper is organized as follows. In \S \ref{sec:sample}, we present details of the observations and data reduction, including a description of the systematic technique used to calculate \lya equivalent widths. Stellar population modeling is discussed in \S \ref{sec:pops}. The properties of objects with and without strong \lya emission are presented in \S \ref{sec: results} and we discuss how our data can be used to address several of the outstanding questions pertaining to the physical nature of LBGs and LAEs in \S \ref{sec: disc}. A summary and our conclusions appear in \S \ref{sec: sum}. We assume a standard $\Lambda$CDM cosmology throughout with \emph{H}$_{\rm 0}$ = 70 \kms\ Mpc$^{-1}$, $\Omega_{\rm M}$ = 0.3, and $\Omega_{\Lambda}$ = 0.7. All magnitudes are based on the AB system \citep{oke1983}\footnote{AB magnitude and $f_{\nu}$, the flux density in units of ergs s$^{-1}$ cm$^{-2}$ Hz$^{-1}$, are related by \emph{m}$_{\rm AB}$ = $-$2.5~log$_{10}~f_{\nu}-$48.6. Conversions between AB and Vega magnitudes for the near-infrared passbands are as follows: $K_s$(AB) = $K_s$(Vega) + 1.82; $J$(AB) = $J$(Vega) + 0.90.}, with the exception of the infrared passbands which are in the Vega system. 

\section{Observations and Data Reduction}  \label{sec:sample} 
\subsection{Imaging and Spectroscopy}
The data presented here are drawn from the LBG surveys of Steidel and collaborators, with approximately half of the observations described in \citet{steidel2003, steidel2004} and half from subsequent programs by the same authors. These surveys employed photometric preselection in the $U_nG{\cal R}$ passbands in a variety of fields \citep{reddy2008} to target galaxies in the redshift interval $z$ $\sim$ 2--3. Follow-up optical spectroscopy of a subset of these galaxies, paired with supplemental near and mid-infrared photometry, has yielded an extensive data set upon which multiple studies have been based \citep[e.g.,][]{shapley2003,adelberger2005c,shapley2005a,erb2006b,reddy2008}.

Here, we introduce a spectroscopic and photometric sample of \z3 LBGs. These data were photometrically preselected with the following standard LBG $U_nG{\cal R}$ flux and color cuts: \begin{equation} \label{color-cuts} {\cal R}\le 25.5,\quad G-{\cal R}\le 1.2,\quad U_n-G\ge G-{\cal R}+1
\end{equation} where the $U_n$, $G$, and ${\cal R}$ passbands sample $\lambda_{\rm rest}$ $\sim$ 900, 1200, and 1700 \AA\ at \z3, respectively. Object detection, color cuts, and photometry are discussed in \citet{steidel2003}. Multi-object optical spectroscopy was obtained using the Low Resolution Imaging Spectrometer \citep[LRIS;][]{oke1995} on the Keck \Rmnum{1} 10m telescope. The majority of the data (93\%) were taken with the blue arm of LRIS \citep[LRIS-B;][]{mccarthy1998,steidel2004}, and the remainder of the data were obtained with LRIS prior to its blue arm upgrade in September 2000. The LRIS-B data were collected using 300, 400, and 600 line mm$^{-1}$ grisms, which resulted in spectral resolutions of $\lambda/\Delta\lambda$ = 1000, 1200, and 2000, respectively. The 400 (600) line mm$^{-1}$ grism was used for 55\% (39\%) of the observations,  and the remaining $\sim$ 6\% of the LRIS-B spectra were obtained with the 300 line mm$^{-1}$ grism. LRIS-B rest-frame wavelength coverage extended from $\sim$ 900--1500 \AA\ and a typical integration time was 3 $\times$ 1800 s. The data were reduced (flat-fielded, cosmic ray rejected, background subtracted, extracted, wavelength and flux calibrated, and transformed to the vacuum wavelength frame) using {\tt IRAF} scripts. Details of the data collection and reduction of both the preselection and spectroscopic samples are presented in \citet{steidel2003}. 

\begin{deluxetable}{llrll}
\tablewidth{0pc}
\tablewidth{0pt}
\tablecaption{\textsc{Spectroscopic Survey Fields}\label{table: field}}
\tablehead{
\multicolumn{1}{c}{Field Name}
& \multicolumn{1}{c}{{$\alpha$}\tablenotemark{a}}
& \multicolumn{1}{c}{{$\delta$}\tablenotemark{b}}
& \multicolumn{1}{c}{Field Size}
& \multicolumn{1}{c}{N$_{\rm LBG}$\tablenotemark{c}} \\
\colhead{}
& \multicolumn{1}{c}{(J2000.0)}
& \multicolumn{1}{c}{(J2000.0)}
& \multicolumn{1}{c}{(arcmin$^2$)}
& \multicolumn{1}{c}{}
 }
\startdata
Q0100$^{\star}$ & 01 03 11 & 13 16 18 & 42.9 & 22 \\
Q0142 & 01 45 17 & --09 45 09 & 40.1 & 20 \\
Q0449 & 04 52 14 & --16 40 12 & 32.1 & 13 \\
Q1009$^{\star}$ & 10 11 54 & 29 41 34 &  38.3 & 30 \\
Q1217 & 12 19 31 & 49 40 50 & 35.3 &13 \\ 
GOODS-N\tablenotemark{d}$^{\star}$$^{\dagger}$ & 12 36 51 & 62 13 14 & 155.3 & 54 \\ 
Q1307 & 13 07 45 & 29 12 51 & 258.7 & 8 \\
Q1549$^{\star}$$^{\dagger}$ & 15 51 52 & 19 11 03 & 37.3 & 48 \\ 
Q1623$^{\star}$$^{\dagger}$ & 16 25 45 & 26 47 23 & 290.0 & 24 \\
Q1700$^{\star}$$^{\dagger}$ & 17 01 01 & 06 11 58 & 235.3 & 39 \\
Q2206$^{\star}$ & 22 08 53 & --19 44 10 & 40.5 & 23 \\
Q2343$^{\star}$$^{\dagger}$ & 23 46 05 & 12 49 12 & 212.8 & 26 \\
Q2346 & 23 48 23 & 00 27 15 & 280.3 & 1 \\ 
& & & & \\
\textbf{TOTAL} & \textbf{ ...} & \textbf{...} & \textbf{1698.9} & \textbf{321} \\
\enddata
\tablenotetext{a}{Right ascension in hours, minutes, and seconds.}
\tablenotetext{b}{Declination in degrees, arcminutes, and arcseconds.}
\tablenotetext{c}{Number of spectroscopically-confirmed LBGs with redshifts \emph{z} $\ge$ 2.7, excluding QSOs and AGN.}
\tablenotetext{d}{This field is also referred to as ``HDF".}
\tablenotetext{$\star$}{Denotes a field with near-infrared imaging.}
\tablenotetext{$\dagger$}{Denotes a field with mid-infrared \emph{Spitzer} IRAC imaging.}
\end{deluxetable}

Approximately 3\% of the spectroscopically-confirmed \z3 LBGs were classified as either active galactic nuclei (AGN) or quasi-stellar objects (QSOs) on the basis of broad lines and high-ionization emission features, respectively \citep{reddy2008}. These objects were excluded from the spectroscopic sample, as were galaxies at redshifts $z$ $\le$ 2.7. The final sample, spanning 13 photometric preselection fields totaling 1700 arcmin$^2$, includes 321 objects with an average redshift of $\langle z \rangle$ = 2.99 $\pm$ 0.19 (Table \ref{table: field}, Figure \ref{fig: redshift}). We note that this sample is distinct from previous studies of \z3 LBGs \citep[e.g.,][]{shapley2001,shapley2003} in that the majority of these objects have corresponding near- and mid-infrared photometry. 

Near-infrared photometry in the $J$ ($\lambda_{\rm c}$ = 1.25 $\mu$m) and $K_s$ ($\lambda_{\rm c}$ = 2.15 $\mu$m) bands was obtained for a subset of the sample (8/13 fields) using the Wide Field Infrared Camera \citep{wilson2003} on the Palomar 5m telescope. 102/321 objects (32\%) were detected in $K_s$ imaging and an additional 69 objects fell on the $K_s$ images and were not detected. We assigned $K_s$ upper limits corresponding to 3$\sigma$ image depths \citep[$K_s$ $\sim$ 22.2 (Vega);][]{erb2006b} to these 69 galaxies. $J$ band photometry was also obtained for 57/102 objects (56\%) detected in the $K_s$ sample. Details of the data collection and reduction of the near-infrared sample are presented in \citet{shapley2005a} and \citet{erb2006b}. 

Mid-infrared imaging was obtained for 5/13 fields with IRAC on \emph{Spitzer}. Observations at 3.6, 4.5, 5.8, and 8.0 $\mu$m were obtained for the GOODS-N \citep{dickinson2002,giavalisco2004,reddy2006b}, Q1700 \citep{shapley2005a}, and Q1549, Q1623, and Q2343 (Erb et al. in preparation) fields, where 3$\sigma$ IRAC detection limits ranged from 25.1--24.8 (AB). The mid-infrared data were reduced according to procedures described in \citet{shapley2005a}. 112/321 objects (35\%) have detections in at least one IRAC passband, and 34/321 objects (11\%) have both $K_s$ and IRAC detections. 

\subsection{Galaxy Systemic Redshifts}

In order to prepare the spectra for subsequent measurement and analysis, we transformed each spectrum into the stellar systemic frame where the galaxy's center of mass was at rest. To do so, we employed the procedure of \citet{adelberger2003} to infer the galaxy's systemic redshift from measurements of its redshifts of both \lya in emission and interstellar lines in absorption. For the spectra that clearly exhibited a double-peaked \lya emission feature (12/321 objects), we adopted the convention of setting the \lya emission redshift equal to the average redshift of the two emission peaks. This technique of inferring a zero-velocity center-of-mass redshift, as opposed to measuring it directly, was necessary due to the fact that stellar lines arising from OB stars (assumed to be at rest with respect to the galaxy) are too weak to measure in individual spectra at \z3. Furthermore, a systemic redshift could not be measured from prominent LBG spectral signposts (e.g., \lya or interstellar absorption lines) as these features trace outflowing gas which is offset from the galaxy's center-of-mass frame by several hundred \kms\ \citep{shapley2003}. 

\subsection{\lya Equivalent Width} \label{sec: EW}

H\Rmnum{1} Ly$\alpha$, typically the strongest feature in LBG spectra, is characterized by its equivalent width, \ewlya, where we use a negative equivalent width to correspond to the feature in absorption. We present here a systematic method for estimating \ewlya, taking into account the various \lya spectroscopic morphologies that were observed in the sample. In particular, this method employs a more robust technique than used previously to determine the wavelength extent over which the \lya feature should be integrated to extract a line flux. 

We first binned the 321 systemic-frame spectra into one of four categories based on the morphology of Ly$\alpha$: ``emission", ``absorption", ``combination", and ``noise". The spectra in the ``emission" bin were clearly dominated by a \lya emission feature, and a small subset of this sample exhibited two peaks in emission. The spectra in the ``absorption" bin were dominated by a trough around Ly$\alpha$, typically extending for tens of angstroms bluewards of line center. The spectra deemed to be ``combination" contained a \lya emission feature superimposed on a larger \lya absorption trough and the ``noise" spectra were generally featureless around Ly$\alpha$, save for a possible absorption signature whose secure identification was hindered by low signal-to-noise. Four example spectra, characterized as falling into each of these four bins, are shown in Figure {\ref{fig: categories}.

\begin{figure}[t]
\centering
\includegraphics[width=3in]{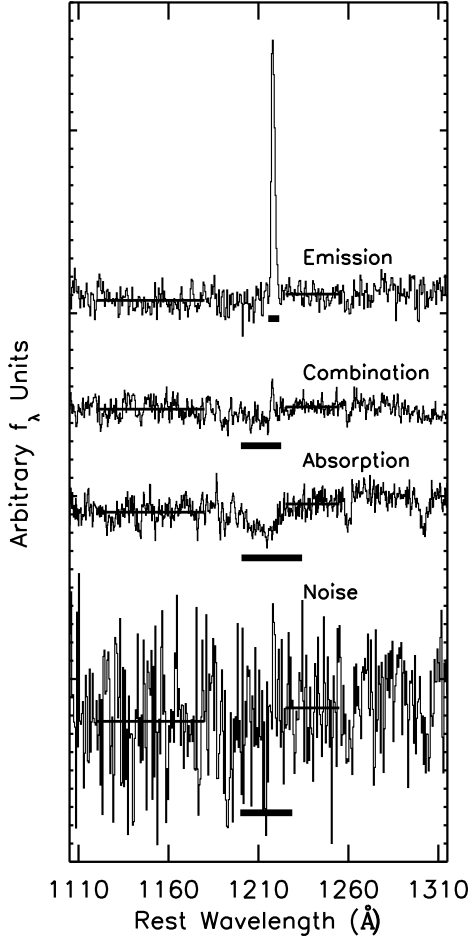}
\caption{The \lya feature varies widely in its morphology. Four spectra are plotted to show representative examples of objects classified in the ``emission," ``combination," ``absorption," and ``noise" bins, respectively. In order to systematically calculate \lya equivalent width, we adopted red- and blue-side continua (horizontal lines from 1120--1180 \AA\ and from 1225--1255 \AA, respectively) and inferred the extent of the \lya feature (thick line below each spectrum) using the methodology described in \S \ref{sec: EW}. Note that in the case of the ``absorption" spectrum shown here, the extent of the \lya feature appears to extend redwards of the adopted red-side continuum -- this difference arises because the plotted spectrum is unsmoothed while a smoothed spectrum was employed to calculate the wavelength bounds of the \lya feature.}
\label{fig: categories} 
\end{figure}

Each spectrum, regardless of its category classification, was fit with two average continuum levels, one bluewards  (1120--1180 \AA; c$_{\rm blue}$) and one redwards (1225--1255 \AA; c$_{\rm red}$) of Ly$\alpha$; these wavelength ranges were chosen to avoid the prominent Si \Rmnum{3} and Si \Rmnum{2} absorption features at 1206 and 1260 \AA, respectively. We worked with both the spectra and the adopted continua in $f_{\lambda}$ units (erg s$^{-1}$ cm$^{-2}$ \AA$^{-1}$). Below, we briefly describe the procedure for calculating \ewlya\ for each of the four morphological classification bins.

\emph{Emission:} 189/321 objects (59\%):  The wavelength of the maximum flux value between 1213--1221 \AA\ was calculated, as well as the wavelengths on either side of the maximum where the flux level intersected c$_{\rm red}$ and c$_{\rm blue}$, respectively. These latter two wavelengths were adopted as the extremes of the emission feature. In a limited number of cases (12/188 objects), double-peaked spectra were individually examined to ensure that this methodology counted both peaks as contained within the \lya feature. The {\tt IRAF} routine {\tt SPLOT} was next used to calculate the enclosed flux between the two wavelength bounds. The enclosed flux was then divided by the level of c$_{\rm red}$ to yield a measurement of \ewlya\ in \AA. The level of c$_{\rm blue}$ was not used in the calculation of \ewlya\ due to its substantial diminution by the intergalactic medium (IGM).

\emph{Absorption:} 50/321 objects (16\%): The boundaries of the \lya absorption feature were calculated in the same manner as those of the ``emission" spectra described above, with the exception that the flux value between 1213 and 1221 \AA\ was isolated as a minimum and the ``absorption" spectra were initially smoothed with a boxcar function of width six pixels ($\sim$ 2.5 \AA) in order to minimize the possibility of noise spikes affecting the derived wavelength boundaries of the \lya feature. These smoothed spectra were only used to define the extent of the \lya line; the original unsmoothed spectra were used for the flux integration in {\tt IRAF} and the enclosed flux was divided by c$_{\rm red}$ to yield \ewlya.

\emph{Combination:} 31/321 objects (10\%): Objects in the ``combination" bin were characterized by a \lya emission feature superposed on a larger absorption trough. The boundaries of the \lya feature were computed by beginning at the base of the \lya emission peak and moving toward larger fluxes until the smoothed spectrum (see above) intersected $\rm c_{\rm red}$ and c$_{\rm blue}$, respectively (the same technique used for the ``absorption" spectra). Flux integration and division by c$_{\rm red}$ were furthermore identical to those objects discussed above. 

\emph{Noise:} 51/321 objects (16\%): For these spectra dominated by noise, we adopted set values for the endpoints of the \lya feature based on the average boundary values of the absorption and combination spectra -- 1199.9 and 1228.8 \AA. (The boundaries of the ``emission" spectra were not included in this calculation, as the spectral morphologies of the ``emission" galaxies differed greatly from those of the ``noise" galaxies). As above, the integrated flux was divided by the level of c$_{\rm red}$ to yield \ewlya.

Rest frame equivalent widths ranged from $-$40 \AA\ $\lesssim$ \ewlya\ $\lesssim$ 160 \AA, although one object (HDF--C41) had an equivalent width of $\sim$ 740 \AA; we attributed this outlier to a continuum level in the spectrum comparable with zero and omitted this object from further analysis. The median equivalent width of the sample was $\sim$ 4 \AA\ (Figure \ref{fig: EW}), consistent with values reported by \citet{shapley2001,shapley2003} for \z3 LBGs. 

\subsection{Composite Spectra} \label{sec: comp}

A composite spectrum offers the distinct advantage of higher signal-to-noise over individual observations. We accordingly constructed several composite spectra from our sample, using, in each case, the same basic steps discussed below.

\begin{figure}[t]
\centering
\includegraphics[trim=0in 0in 0in 0in,clip,width=3in]{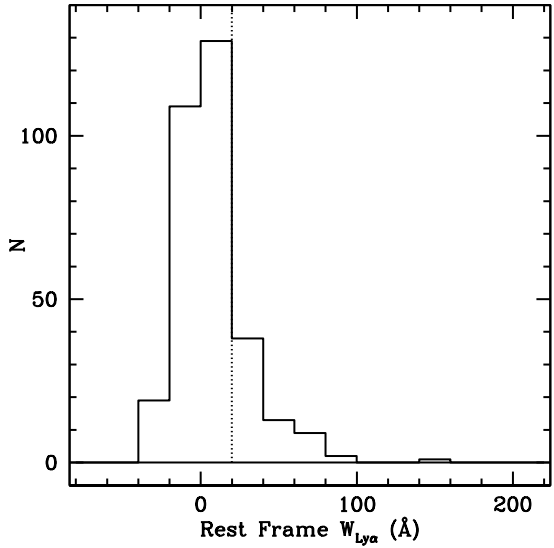}
\caption{The distribution of \lya rest-frame equivalent widths, where objects with \ewlya\ $\ge$ 20 \AA\ (dashed line) were classified as LAEs. The median equivalent width of the sample is $\sim$ 4 \AA, and an outlier at $\sim$ 740 \AA\ (attributed to an uncertain continuum level) is not shown for visual clarity.}
\label{fig: EW} 
\end{figure}

The one-dimensional, flux-calibrated, rest-frame input spectra of interest were stacked (mean-combined) using the {\tt IRAF} {\tt scombine} routine. Each input spectrum was scaled to a common mode over the wavelength range 1250--1380 \AA\ and a small number of positive and negative outliers ($<$ 10\% of the data) were rejected at each pixel position to prevent poor sky-subtraction or cosmic ray residuals from affecting the composite spectrum. The final composite spectrum was then rebinned to a dispersion of 1 \AA\ pixel$^{-1}$. A composite spectrum\footnote{This composite spectrum is meant to represent only the average of the objects in our sample, not the average of the entire \z3 LBG population. There are a variety of observational biases that affect that relative proportions of Ly$\alpha$-emitters and Ly$\alpha$-absorbers selected: large \lya emission lines contaminating the $G$ band result in redder $U_n$ -- $G$ colors, scattering objects into the color section window (Equation \ref{color-cuts}), while \lya absorption limits the dynamic range in continuum magnitude over which objects are selected. We refer the reader to \citet{steidel2003} for a detailed discussion of these biases.} of the entire sample is shown in Figure \ref{fig: global}, where several photospheric and low- and high-ionization interstellar features are visible in addition to Ly$\alpha$.

\section{Stellar Population Modeling}  \label{sec:pops} 

As the majority of the spectroscopic sample had extensive accompanying photometric wavelength coverage, we conducted population synthesis modeling to derive the key properties of age, extinction, star formation rate, and stellar mass. We required at least one photometric measurement redwards of the Balmer break ($\lambda_{\rm rest}$ = 3646 \AA) for robust population synthesis modeling. 248/320 objects satisfied this criterion of photometry in at least one near-infrared or IRAC passband, including 69 objects with $K_s$ upper limits (and no IRAC data).  We modeled galaxies using \citet{bc2003} SEDs (assuming a \citet{salpeter1955} initial mass function (IMF) over the mass range 0.1--125 $M_{\odot}$) and the Calzetti (2000) \nocite{calzetti2000} extinction law derived from local starbursts, where dust extinction, parameterized by E(B--V), was estimated from the latter. While the \citet{calzetti2000} law appears valid on average for \z3 LBGs, we discuss in \S \ref{sec: bootstrap} some caveats associated with adopting this law; we also present in that same section a brief discussion of our adoption of the \citet{bc2003} population synthesis models. We furthermore assumed a constant star formation history and solar metallicity. Recent work has suggested that the \citet{salpeter1955} IMF has too steep a slope below 1 $M_{\odot}$ and consequently overpredicts the mass-to-light ratio and stellar mass by a factor $\sim$ 2 \citep{bell2003,renzini2006}. We accordingly converted stellar masses and SFRs to the \citet{chabrier2003} IMF by dividing the model output values by 1.8. The modeling procedure is described in detail in \citet{shapley2001}; we briefly present a summary below. 

Firstly, for the subset of the sample where \lya fell in the $G$ bandpass (2.48 $\lesssim$ $z$ $\lesssim$ 3.38; 238/248 objects), we corrected the $G$ band photometry to account for the discrepancy between the equivalent widths of the sample ($-$40 \AA\ $\lesssim$ \ewlya\ $\lesssim$ 160 \AA) and those of the \citet{bc2003} model SEDs (\ewlya\ $\sim$ $-$10 \AA). According to the formula in \citet{papovich2001}, we applied a correction in the cases where the incremental change in \emph{G} magnitude, $\Delta \emph{G}$, was larger than the uncertainty on the original photometric measurement. This procedure affected 51/238 objects (21\%), for which $\langle \Delta$\emph{G}$\rangle_{\rm median}$ = 0.15 magnitudes. We did not correct for possible contamination in the $K_s$ band from nebular line emission ([O\Rmnum{3}] $\lambda \lambda$5007,4959, H$\beta$ $\lambda$4861), although we tested that our reported correlations (\S \ref{sec: results}) were still robust when all objects at $z$ $\ge$ 2.974 (such that [O\Rmnum{3}] $\lambda$5007, the strongest of these nebular lines,  is shifted into the $K_s$ band) were excluded from the analysis. 

\begin{figure*}[t]
\centering
\includegraphics[trim=0in 0in 0in 0in,clip,width=6in]{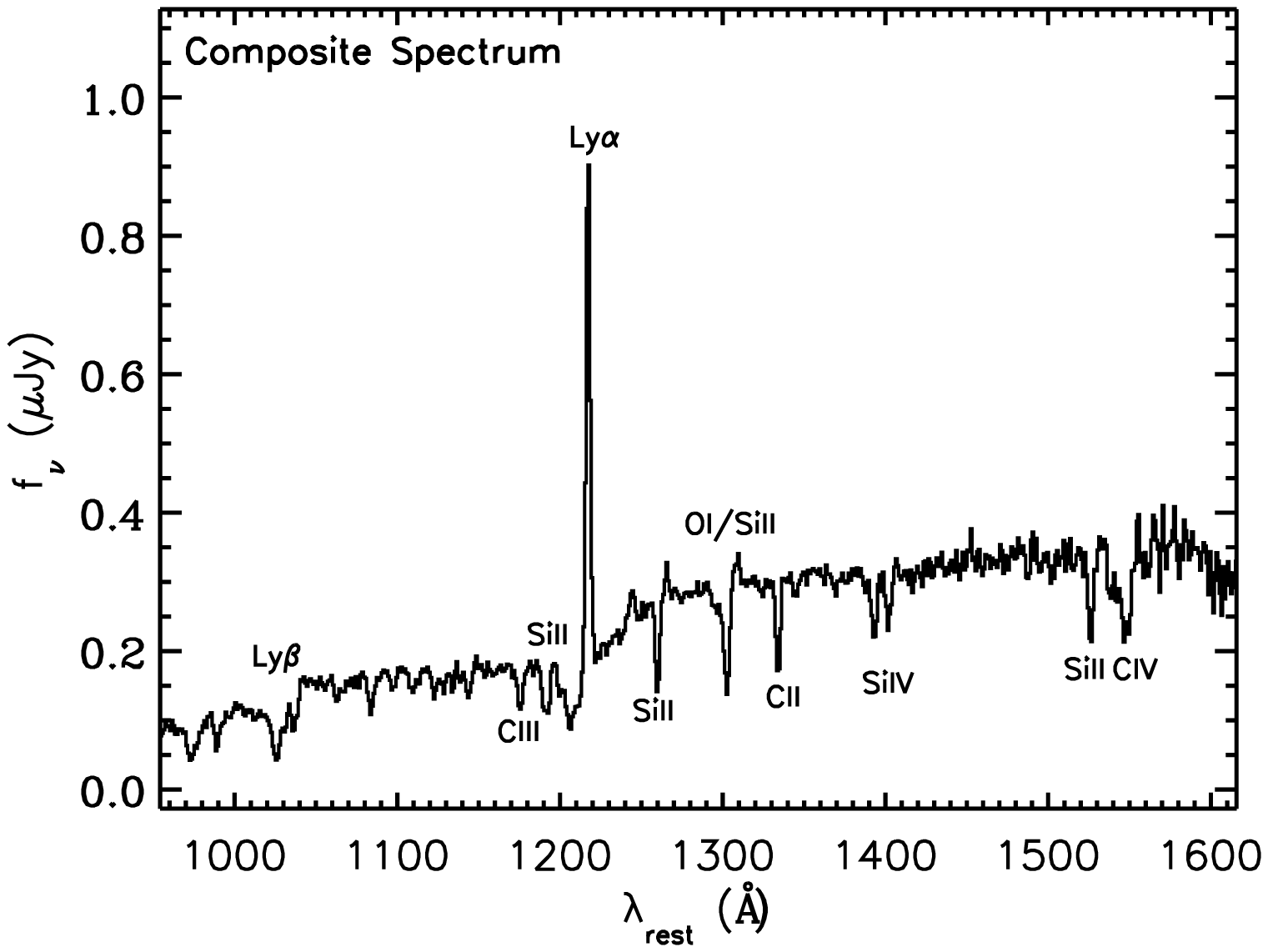}
\caption{Composite rest-frame spectrum assembled from one-dimensional, flux-calibrated spectra (\S \ref{sec: comp}). \lya appears in strong emission and Ly$\beta$ is prominent at 1026 \AA. Photospheric features (e.g., C\Rmnum{3}  $\lambda$1176) and both low- (e.g., Si\Rmnum{2} $\lambda \lambda$1190,1193, Si\Rmnum{2} $\lambda$1260, O\Rmnum{1}+Si\Rmnum{2} $\lambda$1303, C\Rmnum{2} $\lambda$1334, and Si\Rmnum{2} $\lambda$1527) and high- (e.g., Si\Rmnum{4} $\lambda \lambda$1393,1402 and C\Rmnum{4} $\lambda$1549) ionization absorption lines are visible in this high signal-to-noise composite.}
\label{fig: global} 
\end{figure*}

Next, for each galaxy modeled, a grid of SEDs was attenuated by dust and shifted to match the redshift of the galaxy. These redshifted SEDs were further attenuated by IGM absorption \citep{madau1995} and were multiplied by the $G, {\cal R}, J, K_s$, and four IRAC channel filter transmission curves to extract model colors. A model $U_n - G$ color was not calculated, as many of the objects had only upper limits in $U_n$ due to the significant flux diminution in that passband from both galactic and intergalactic H\Rmnum{1} absorption. These predicted colors were then compared with the observed colors by means of the $\chi^2$ statistic and a best-fit E(B--V) and age ($t_{\star}$) were extracted based on the lowest value of $\chi^2$. We also applied the constraint that $t_{\star}$ had to be less than the age of the Universe at the redshift of each object. A best-fit dust-corrected SFR was inferred from the normalization of a galaxy forming stars at a rate of 1 $M_{\odot}$ yr$^{-1}$. Stellar mass, $m_{\rm star}$, was defined as the integral of the SFR and the age of the galaxy. We did not correct stellar masses for interstellar medium (ISM) recycling, whereby a mass fraction of stellar material is returned to the ISM via winds and supernovae \citep{cole2000}. Given that we are concerned with the relative stellar populations of objects modeled in an identical fashion, the constant mass factor introduced by assuming ISM recycling is unimportant. 
 
Best-fit SFR, stellar mass, E(B--V), and age values were extracted for the 179 galaxies without photometric upper limits. The additional 69 galaxies undetected in $K_s$ imaging were modeled by adopting 3$\sigma$ upper limits in ${\cal R}$ -- $K_s$ color as photometric data points. We tested that an upper limit in ${\cal R} - K_s$ color was a robust proxy for an upper limit in both age and stellar mass by perturbing the ${\cal R} - K_s$ colors in several increments, both redwards and bluewards, and re-modeling the perturbed SEDs (holding all other colors constant). The expected trend that a redder galaxy would be best fit with an older age and a larger stellar mass was borne out for the entire sample. We additionally found that E(B--V) and SFR were generally insensitive to perturbations in ${\cal R} - K_s$ color. Therefore, given a galaxy with an upper limit in ${\cal R}$ -- $K_s$ color, its model age and stellar mass were adopted as upper limits and its model E(B--V) and SFR were assumed to be best-fit values. 

Considering the entire sample of 248 objects, we found median best-fit SFR, stellar mass, E(B--V), and age values of 37 $M_{\odot}$ yr$^{-1}$, 7.2 $\times$ 10$^{9}$ $M_{\odot}$, 0.170, and 320 Myr, respectively. When objects with $K_s$ upper limits (and corresponding upper limits in stellar age and mass) were removed from the analysis, the medians were 51 $M_{\odot}$ yr$^{-1}$, 8.3 $\times$ 10$^{9}$ $M_{\odot}$, 0.180, and 320 Myr, respectively. A dust attenuation described by E(B--V) = 0.170 corresponds to $A_V$ $\sim$ 0.7 magnitudes, using the \citet{calzetti2000} starburst attenuation law with ${R'_V}$ = 4.05.  Histograms of the best-fit values are shown in Figure \ref{fig: histogram}.

For each galaxy, confidence intervals of the best-fit stellar parameters were estimated with Monte Carlo simulations. First, each object's photometry was perturbed by an amount drawn from a Gaussian distribution described by the photometric error. Then, the galaxy was remodeled with these new ``observed" colors, and this process was repeated 1000 times. These simulations resulted in estimates of the distributions of best-fit stellar parameters allowed by the uncertainties in the photometry. In some instances, these confidence intervals were not centered on the best-fit stellar population parameters derived from fitting the photometry. In all cases, we proceeded to adopt the best-fit values as representative of each galaxy's properties, and we furthermore assumed that the error on each parameter was described by the standard deviation of that parameter's confidence interval. We used the quantity $\sigma_{\rm x}$/$\langle x \rangle$ to express the ratio of the standard deviation of each parameter's confidence interval, $\sigma_{\rm x}$, to the mean of that parameter's confidence interval, $\langle x \rangle$. On average, using 3$\sigma$ rejection to suppress the effect of outliers, we found that the median values of $\sigma_{\rm x}$/$\langle x \rangle$ for each of the four best-fit stellar population parameters (E(B--V), age, SFR, and stellar mass) were $\sim$ 0.4, 1.0, 0.8, and 0.6, respectively.

\begin{figure*}[t]
\centering
\includegraphics[width=6.5in]{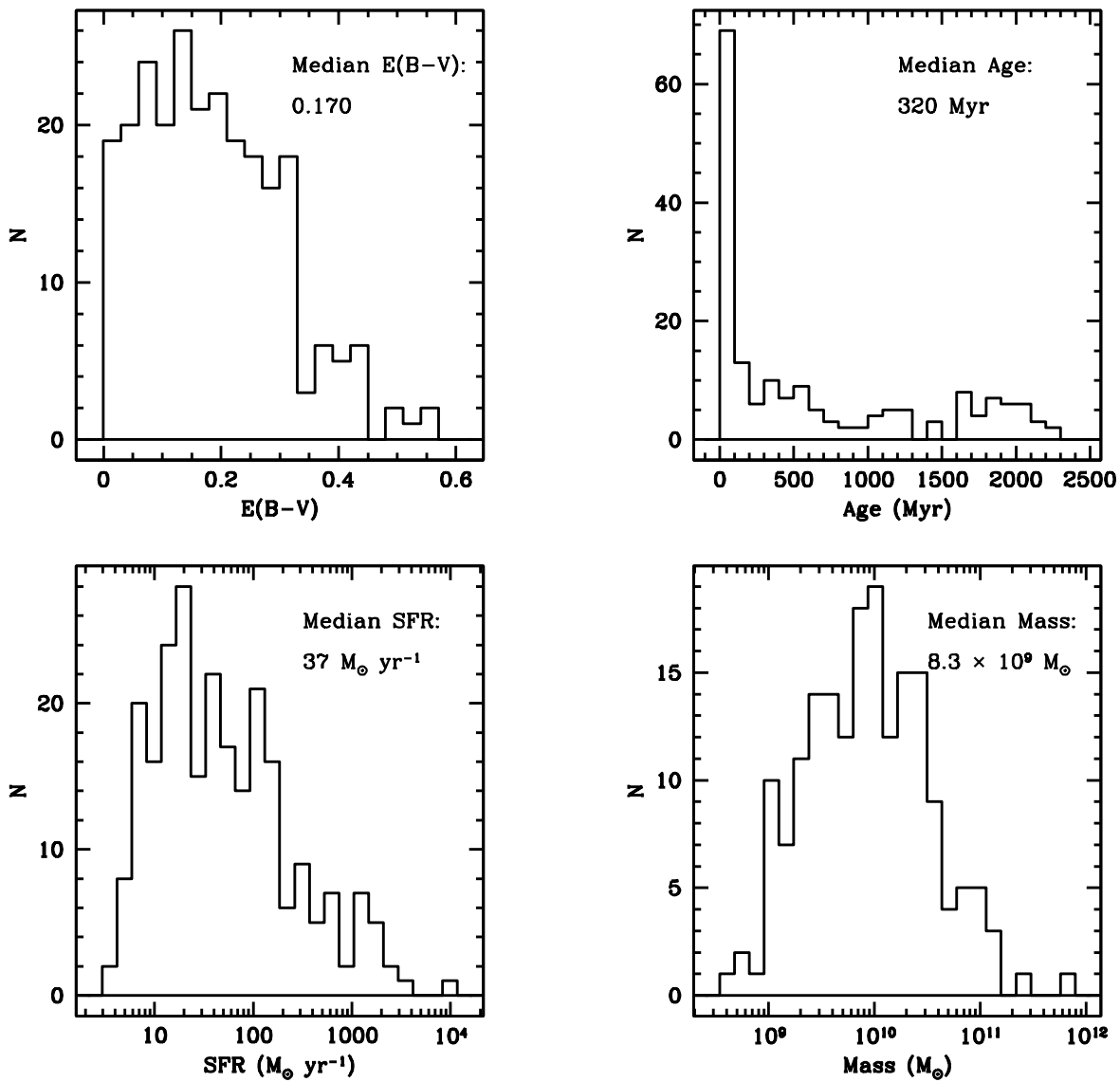}
\caption{Histograms of best-fit stellar population parameters, with median values indicated. $K_s$-undetected galaxies were omitted when making the age and stellar mass distributions, as these parameters represent upper limits in the presence of a $K_s$ non-detection. Dust attenuations span the range E(B--V) = 0.0--0.6, where the median extinction level corresponds to $A_{\rm 1600}$ ($A_V$) $\sim$ 1.7 (0.7) magnitudes. While LBGs are fit with a variety of ages, note the large number of objects in the youngest age bin, $t_{\star}$ $<$ 100 Myr. The SFRs of LBGs fall between those of quiescent, Milky Way-type galaxies \citep[SFR $\sim$ 4 $M_{\odot}$ yr$^{-1}$; e.g.,][]{diehl2006} and vigorously star-forming submillimeter galaxies \citep[SFR $\sim$ 1000 $M_{\odot}$ yr$^{-1}$; e.g.,][]{chapman2005}. Stellar masses range from $\sim$ 10$^9$ to 10$^{11}$ $M_{\odot}$, where the highest mass object, 6 $\times$ 10$^{11}$ $M_{\odot}$, is extremely red (${\cal R}$ -- $K_s$ = 4.48).}
\label{fig: histogram} 
\end{figure*}

\section{Stellar Populations \& \lya Line Strength} \label{sec: results}

With \lya equivalent widths and best-fit stellar population parameters in hand, we now turn to examining the relationship between \lya emission and stellar populations in LBGs. The aim of this analysis is to investigate the physical nature of \z3 LBGs by studying how objects with and without strong \lya emission differ in the fundamental parameters of age, stellar mass, extinction, and star formation rate. Our full complement of data are used in this analysis, including rest-frame UV spectroscopy, broadband photometry, and best-fit stellar population parameters and SEDs. We also make comparisons between our LBG data and narrowband-selected LAEs from \citet{nilsson2007} and \citet{gawiser2007}, although we caution that the relationship between LBGs and narrowband-selected LAEs is a field of extragalactic research unto itself; we refer the reader to \S \ref{sec: caveat} for a summary of the salient points of this topic. 

\subsection{Statistical Tests}

We employed several statistical methods central to our investigation, including survival analysis techniques that were capable of analyzing data with limits (``censored data"). It was important to include limits in the statistical analysis as a non-negligible fraction of the sample (28\%) was undetected in $K_s$ imaging. We used ASURV (``\emph{A}stronomy \emph{SURV}ival \emph{A}nalysis") Rev 1.2 \citep{isobe1990,lavalley1992}, which implements the methods presented in \citet{feigelson1985} and \citet{isobe1986}. The bivariate correlation tests Kendall $\tau$ and Spearman $\rho$ were utilized, where, for each test, the degree of correlation is expressed in two variables: $\tau_{\rm K}$ ($r_{\rm SR}$) and $P_{\rm K}$ ($P_{\rm SR}$). The first variable represents the test statistic and the second variable represents the probability of a null hypothesis (i.e., the probability that the data are uncorrelated). We also used routines from Numerical Recipes \citep{press1992}, specifically the Kolmogorov-Smirnov (K--S) test between two data sets ({\tt kstwo}), the Spearman's rank correlation ({\tt spear}), and the Kendall's $\tau$ correlation ({\tt kendl1}). The K--S test produces two outputs: $D$ and $P$, where the former is the test statistic and the latter is the probability that the two samples are drawn from the same underlying population. {\tt spear} and {\tt kendl1} are duplicates of the Spearman $\rho$ and Kendall $\tau$ tests available in ASURV; we employed the ASURV routines when it was necessary to include data with limits. 

\subsection{Equivalent Width Versus Stellar Parameters} \label{sec: bootstrap}

Firstly, we investigated the correlation between \lya equivalent width and extinction, age, star formation rate, and stellar mass, respectively, using the Spearman $\rho$ and Kendall $\tau$ bivariate correlation tests. For this analysis, we used the entire sample with population modeling, including galaxies with upper limits in the $K_s$ passband (248 objects). We note that while the correlations shown in Figure \ref{fig: 1234} are not without scatter, we have employed a variety of quantitative statistical analyses that have yielded a consistent picture of how stellar populations are correlated with \lya emission in \z3 LBGs. Below, we introduce these results.

\begin{figure*}[t]
\centering
\includegraphics[width=7in]{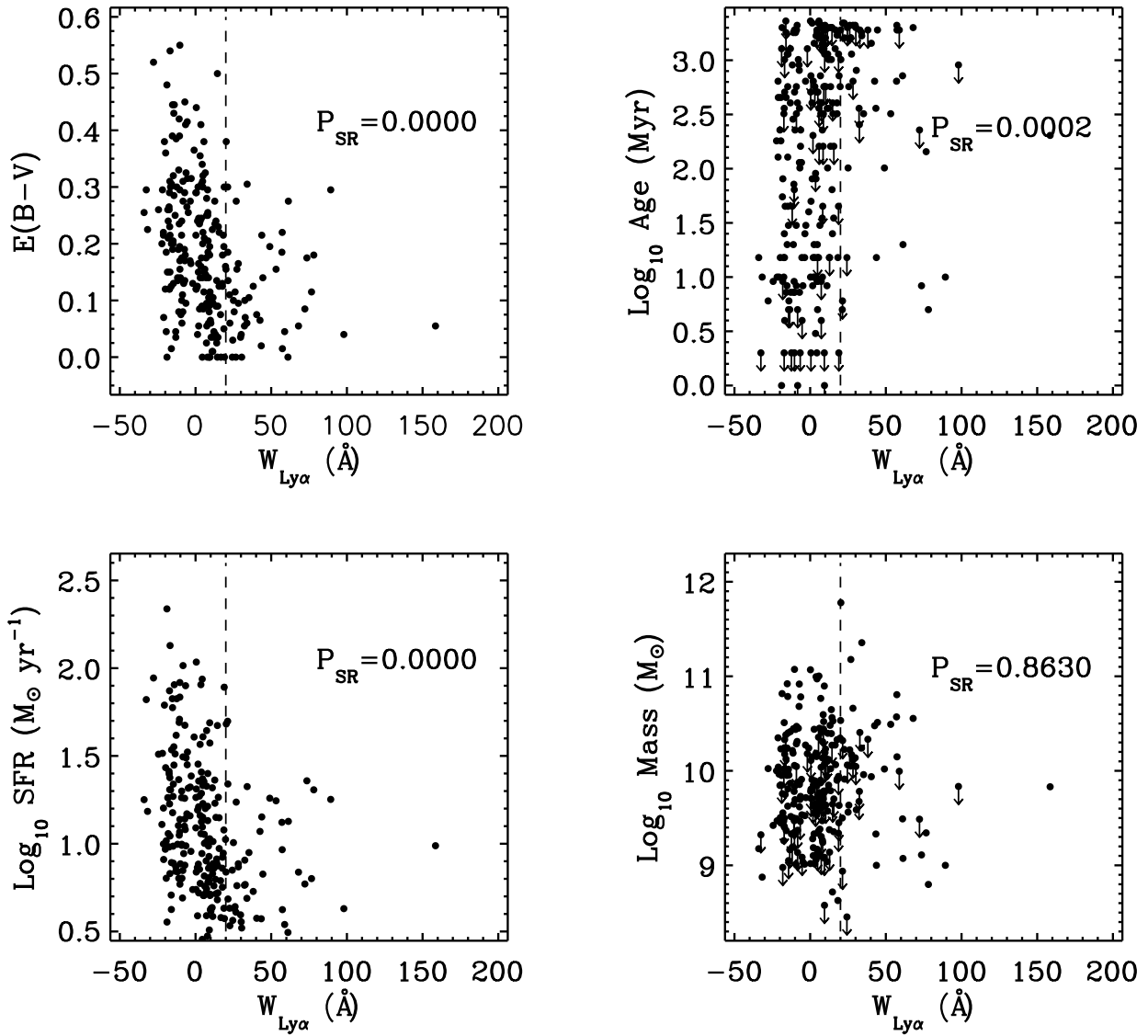}
\caption{Plots indicating the correlation between \ewlya\ and each of the four best-fit stellar population parameters. Upper limits in age and mass are indicated by downward-facing arrows and the vertical dashed line in each panel delineates \ewlya\ = 20 \AA, the adopted lower bound in \lya equivalent width for LAEs. The probability of a null hypothesis (i.e., no correlation) from the Spearman $\rho$ test is shown in each plot. Note the significant correlation between equivalent width and E(B--V), age, and SFR, respectively, in the sense that galaxies with strong \lya emission also tend to be less dusty, older, and more quiescently forming stars. \ewlya\ and stellar mass are not significantly correlated, as evidenced by the high probability of the null hypothesis.}
\label{fig: 1234} 
\end{figure*}

E(B--V), age, and SFR all exhibited very strong correlations with \ewlya\ ($P_{\rm SR}$ $\leq$ 0.0002 in each case)\footnote{We note, however, that SFR and E(B--V) are not independent; the probability that these two parameters are uncorrelated is $P_{\rm SR}$ = 0.0000.}, whereas stellar mass was uncorrelated ($P_{\rm SR}$ = 0.8630). These results did not change when the 69 galaxies with $K_s$ upper limits (and corresponding upper limits in mass and age) were excluded from the analysis (Table \ref{table: coef}). In order to test the robustness of these results while taking into account the uncertainty on the best-fit parameters, we used the bootstrapping method. We created 100 bootstrap samples by randomly extracting one line from each object's 1000-entry confidence interval of stellar parameters (generated by perturbing the initial photometry according to its errors and then re-modeling the galaxy; \S \ref{sec:pops}) and then re-ran the Spearman $\rho$ test on each of these 100 artificial samples. A null hypothesis was consistently ruled out at between the 3 and 5$\sigma$ levels for the SFR and E(B--V) bootstrap samples and at the 3$\sigma$ level for the age sample. In the case of the stellar mass sample, where the null hypothesis was ruled out at only the 1$\sigma$ level, the data do not support a correlation between \ewlya\ and stellar mass. These bootstrap samples show that even when accounting for the uncertainty in best-fit parameters, a strong correlation exists between equivalent width and age, E(B--V), and SFR, respectively, such that objects with strong \lya emission tend to be older, less dusty, and lower in star formation rate (more quiescent) than objects with weak or no \lya emission.

\begin{deluxetable}{lllr}
\tablewidth{0pc}
\tablewidth{0pt}
\tablecaption{\textsc{Correlation Coefficients}\label{table: coef}}
\tablehead{
  \multicolumn{1}{c}{Parameter}
& \multicolumn{1}{c}{Parameter}
& \multicolumn{1}{c}{Kendall $\tau$}
& \multicolumn{1}{c}{Spearman $\rho$} \\
\colhead{}
& \colhead{}
& \colhead{$\tau_{\rm K}$\tablenotemark{a} ($P_{\rm K}$)\tablenotemark{b}}
& \colhead{$r_{\rm SR}$\tablenotemark{a} ($P_{\rm SR}$)\tablenotemark{b}}
}
\startdata
\cutinhead{Entire Sample (248 objects)}
\ewlya & E(B--V) & 6.585 (0.0000) & --0.413 (0.0000)  \\
..... & Age & 3.655 (0.0003) & ~0.238 (0.0002)  \\
..... & SFR & 6.515 (0.0000) & --0.410 (0.0000)  \\
..... & Stellar Mass & 0.358 (0.7205) & 0.011 (0.8630)  \\
\cutinhead{Omitting $K_s$ non-detections (179 objects)}
\ewlya & E(B--V) & 5.163 (0.0000) & --0.385 (0.0000)  \\
..... & Age & 4.122 (0.0000) & 0.319 (0.0000)  \\
..... & SFR & 5.299 (0.0000) & --0.394 (0.0000)  \\
..... & Stellar Mass & 0.917 (0.3592) & 0.068 (0.3641)  \\
\enddata
\tablenotetext{a}{Test statistic.}
\tablenotetext{b}{Probability of a null hypothesis (i.e., the probability that a correlation is not present).}
\end{deluxetable}

As an independent test of the above results, we next used the rest-frame UV spectra to investigate how the morphology of the \lya feature was, qualitatively, correlated with best-fit stellar population parameters. For this analysis, we limited our study to objects satisfying two criteria: \emph{1)} population synthesis modeling had been completed and \emph{2)} $K_s$ photometry consisted of detections, not upper limits\footnote{Objects with $K_s$ upper limits were excluded due to their corresponding upper limits in best-fit ages and masses; such limits led to uncertainty as to the age and mass bin in which these objects should be grouped.}. Together, these criteria isolated 179 objects. For each stellar population parameter, we divided the objects into three groups (``tertiles") based on their best-fit value. A composite rest-frame UV spectrum was then constructed from each tertile according to the methodology presented in \S \ref{sec: comp}, for a total of 12 composite spectra (3 tertiles $\times$ 4 parameters; Figure \ref{fig: ages}). A strong correlation of equivalent width with age, E(B--V), and SFR was observed, such that stronger \lya emission was more prevalent in older, less dusty, and more quiescent galaxies. At the same time, the strength of low-ionization interstellar absorption lines decreased with increasing \lya emission strength \citep{shapley2003}. We note that these trends were still present when the spectra were median-combined to make to the composites (as opposed to mean-combined). 

\begin{figure*}[t]
\centering
\includegraphics[width=7in]{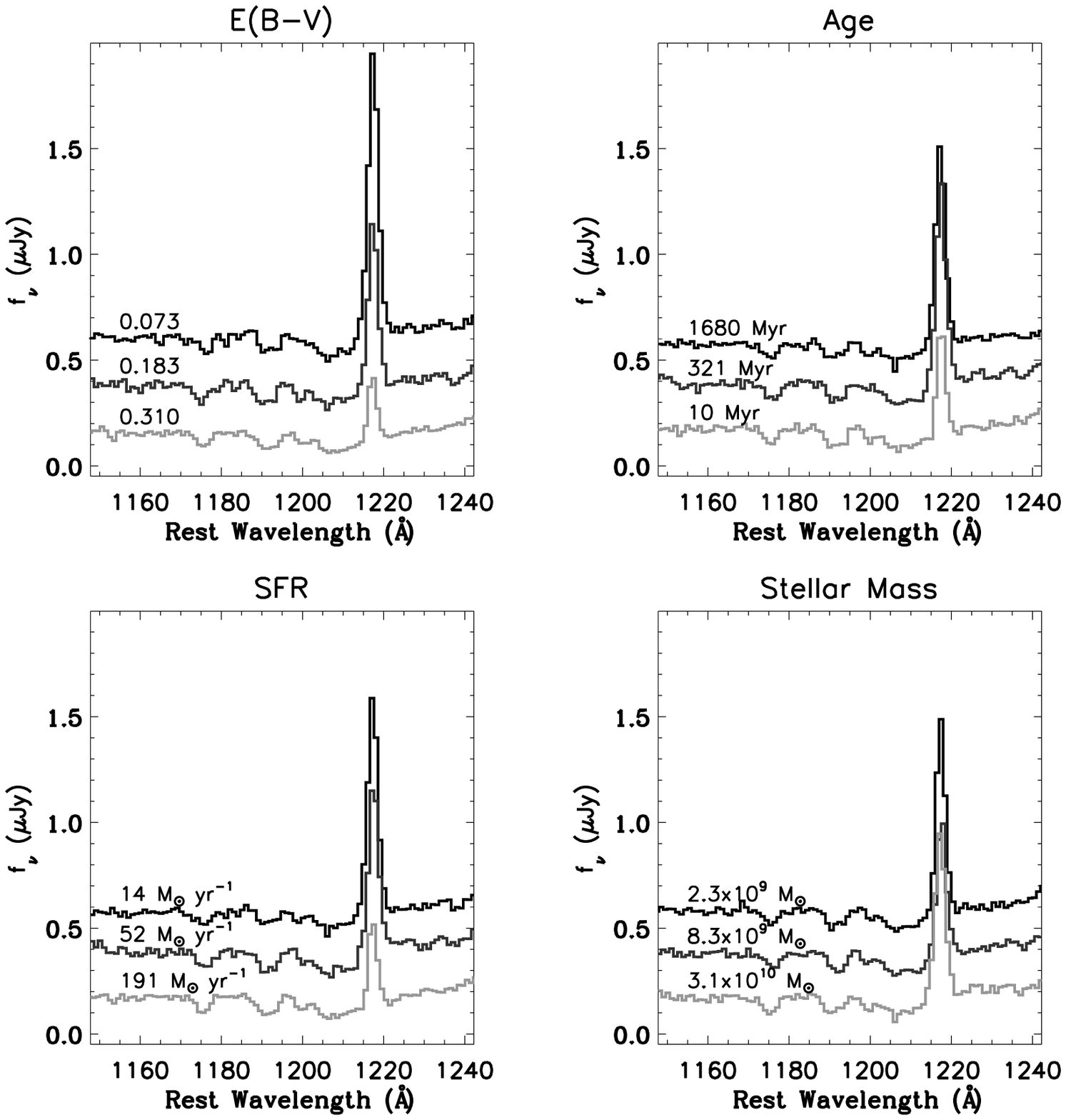}
\caption{Composite spectra assembled from objects ordered by E(B--V), age, SFR, and stellar mass, respectively, where objects with $K_s$ upper limits were omitted. The medium and dark gray spectra are offset by +0.2 and +0.4 $\mu$Jy, respectively, from the light gray spectrum and the median value in each tertile is printed next to its composite. Notice how strong \lya emission is associated with older, less dusty, and more quiescent galaxies. There does not appear to be a significant correlation between \lya emission strength and best-fit stellar mass.}
\label{fig: ages} 
\end{figure*}

Next, we reversed the independent and dependent variables from the previous investigation and, this time, divided the sample into tertiles according to \lya equivalent width. The mean equivalent widths of each tertile were --15.5, 2.8, and 33.1 \AA, respectively. We then calculated the average stellar parameters of each tertile, in addition to the error on the mean (Figure \ref{fig: EW_123}). Consistent with the results of the bivariate correlation tests and the composite spectra divided by stellar population parameters discussed above, we found that as the average equivalent width varied from -15.5 \AA\ to 33.1 \AA, $\langle t_{\star} \rangle$ increased from 440 Myr to 850 Myr, $\langle$E(B--V)$\rangle$ decreased from 0.24 to 0.14, and $\langle$SFR$\rangle$ decreased from 300 $M_{\odot}$ yr$^{-1}$ to 70 $M_{\odot}$ yr$^{-1}$. A weak positive correlation between best-fit stellar mass and equivalent width was observed, although, given the errors on the average masses, we chose to adopt the stance that stellar mass was generally insensitive to changes in equivalent width.

From the three aforementioned investigations, a consistent picture has emerged: strong \lya emission is associated with galaxies that are older, less dusty, and more quiescent than their counterparts with weaker \lya emission or the line in absorption. We note that the trend of increasing \lya strength with decreasing E(B--V) is a well-known result \citep[e.g.,][]{shapley2003,reddy2006a,pentericci2007}. \citet{deharveng2008}, however, found no apparent correlation between \lya equivalent width and UV color in a sample of 96 LAEs at \z0.3, although we caution that the transformation from UV color to E(B--V) is sensitive to the geometry of the emitting stars and absorbing dust and may therefore not be constant for a heterogeneous population such as LAEs \citep[e.g.,][]{witt2000,granato2000}. The relationship between age and \ewlya\ observed in our data -- older objects exhibiting stronger \lya emission than younger objects -- is in agreement with the results of \citet{shapley2003}. The inverse correlation between dust-corrected SFR and \ewlya\ that we observe has been noted by several authors as well \citep[e.g.,][]{ando2004,reddy2006a,tapken2007}, although \citet{nilsson2007} report a $\sim$ 2$\sigma$ direct correlation between SFR and \ewlya\ in a sample of 24 LAEs at \z3.15. In contrast with the results presented here, however, \citet{erb2006a} reported a correlation between stellar mass and \lya strength in a sample of 87 star-forming galaxies at \z2. These authors found that composite spectra constructed from the 30 least massive galaxies ($\langle M \rangle$ = 5 $\times$ 10$^9$ $M_{\odot}$) and the 28 most massive galaxies ($\langle M \rangle$ = 7 $\times$ 10$^{10}$ $M_{\odot}$) in their sample differed widely in nebular line strength, with the former sample exhibiting a pronounced \lya emission feature and the latter sample showing a smaller \lya feature superimposed on a larger absorption trough. \citet{erb2006a} attributed these results to the lower velocity dispersion of the interstellar medium in the less massive galaxies, where, following the arguments of \citet{mashesse2003}, velocity dispersion and \lya escape fraction are anti-correlated. We therefore conclude here that the observed correlations we find between \ewlya\ and E(B--V), SFR, and age are supported by the work of others, whereas the lack of correlation between stellar mass and \ewlya\ may either be a tenable result (and evidence of LBG evolution during the 1.1 Gyr intervening $z$ = 2 and $z$ = 3) or otherwise an indication that further studies of the relationship between \ewlya\ and stellar mass at \z3 are necessary. 

\begin{figure*}[t]
\centering
\includegraphics[width=7in]{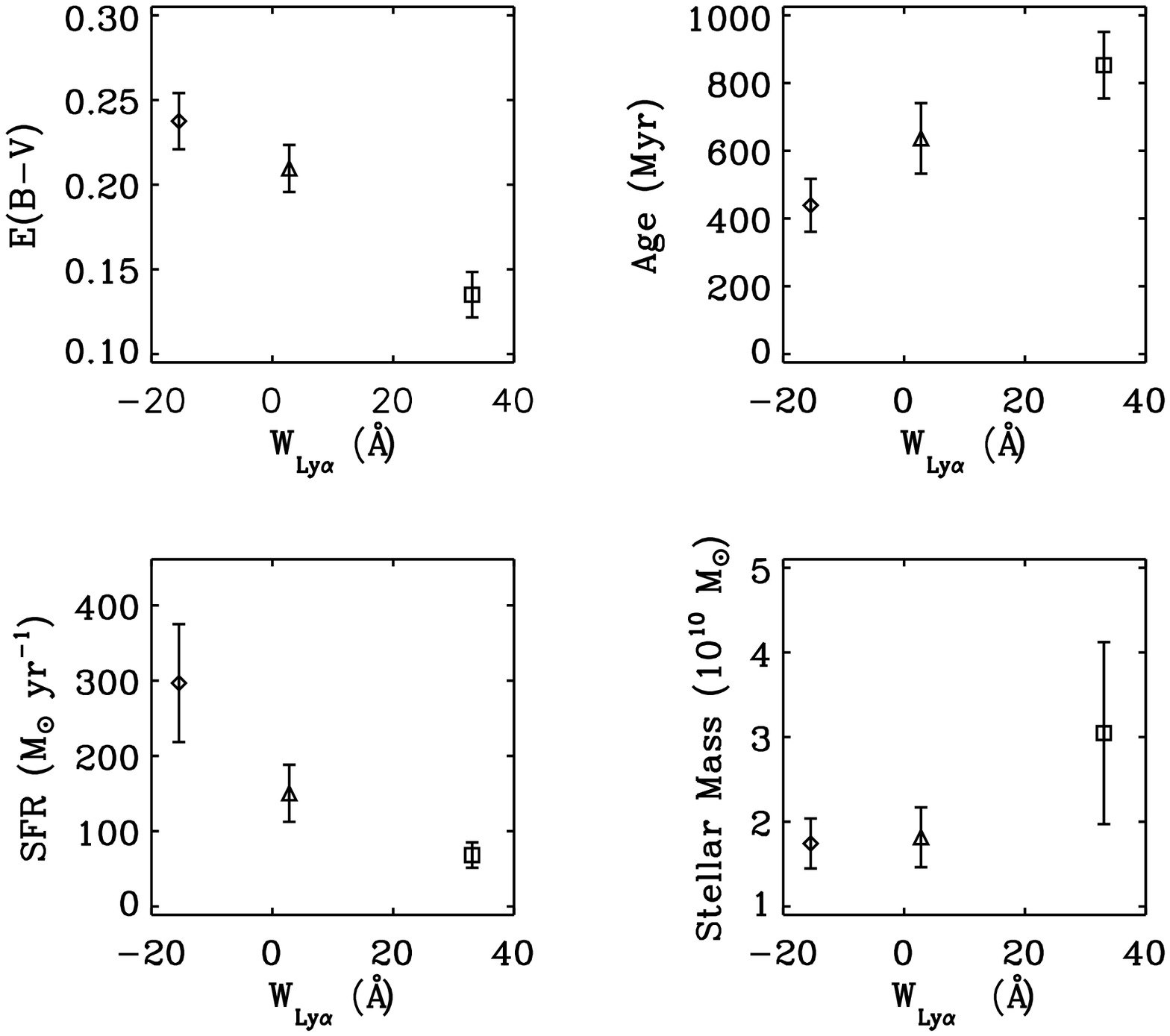}
\caption{Best-fit stellar population parameters versus rest-frame \lya equivalent width. Omitting galaxies with $K_s$ upper limits, the remaining 179 objects with stellar population modeling were ordered by equivalent width and divided into three groups. The average stellar population parameters of each group were extracted, along with the error on the mean ($\sigma$/$\sqrt n$, where $n$ was the number of objects in each group). With increasing \lya strength, galaxies are, on average, older, less dusty, and more quiescent. A weak trend of increasing stellar mass with increasing \lya strength is observed, although this is likely not a significant correlation.}
\label{fig: EW_123} 
\end{figure*}

The correlations discussed above were derived assuming the \citet{calzetti2000} dust attenuation law. While this relation has routinely been applied to samples of high-redshift galaxies \citep[e.g.,][]{erb2006b,gawiser2006,gronwall2007}, recent work by \citet{reddy2006a}, based on \emph{Spitzer} MIPS observations, has suggested that dust extinction in young ($t_{\star}$ $\le$ 100 Myr), UV-selected galaxies at \z2 is often overestimated by modeling a given rest-frame UV color with the \citet{calzetti2000} attenuation law. At \z3, there have been no corresponding statistical studies, but \citet{siana2008,siana2009} have presented preliminary evidence for a similar trend based on two strongly lensed objects. On the other hand, for a single unlensed LBG at \z2.8, \citet{chapman2009} find good agreement between the dust extinction inferred from the \citet{calzetti2000} law and that estimated from the ratio of SCUBA 850 $\mu$m and rest-frame UV fluxes. Based on their findings, \citet{reddy2006a} and \citet{siana2008} accordingly proposed that a steeper, SMC-like extinction relation \citep[e.g.,][]{prevot1984} may be more appropriate for young objects. 

With these results in mind, we assessed the uncertainties inherent in the choice of attenuation law. Limiting our analysis to the 179 objects without $K_s$ upper limits, we re-modeled the 70 galaxies with  $t_{\star}$ $\leq$ 100 Myr with a SMC-like extinction law (and retained the original fits assuming the \citet{calzetti2000} law for the remaining 109 objects). In this mixed sample with some objects modeled with a SMC-like law and others modeled with the \citet{calzetti2000} relation, we found no significant correlations between \ewlya\ and stellar populations. We also investigated the effect of re-modeling all 179 galaxies, regardless of age, with a SMC-like attenuation law. In this case, we observed correlations between \ewlya\ and E(B--V) and between \ewlya\ and SFR. Given the uncertainties in adopting the SMC law for part or all of our sample, another approach consists of restricting our analysis to objects for which the validity of the \citet{calzetti2000} law has not been questioned. When we omitted objects younger than $t_{\star}$ = 100 Myr from our original analysis assuming the \citet{calzetti2000} dust attenuation law, the trends between \ewlya\ and E(B--V), age, and SFR, respectively, were still present (but only at a 2$\sigma$ level of significance). 

Given the lack of systematic studies of the dust attenuation law in \z3 objects, and the uncertainty regarding the age at which a galaxy may transition from being fit with a SMC-like law to the \citet{calzetti2000} law, we choose to present all analysis on results derived from assuming the \citet{calzetti2000} attenuation law. We acknowledge that the choice of an extinction law is a systematic uncertainty in our analysis, yet the persistence of trends between \lya and stellar populations in the sample limited to objects with $t_{\star}$ $\ge$ 100 Myr supports the analysis with the Calzetti law applied to the full sample. 

\begin{figure*}[t]
\centering
\includegraphics[width=7in]{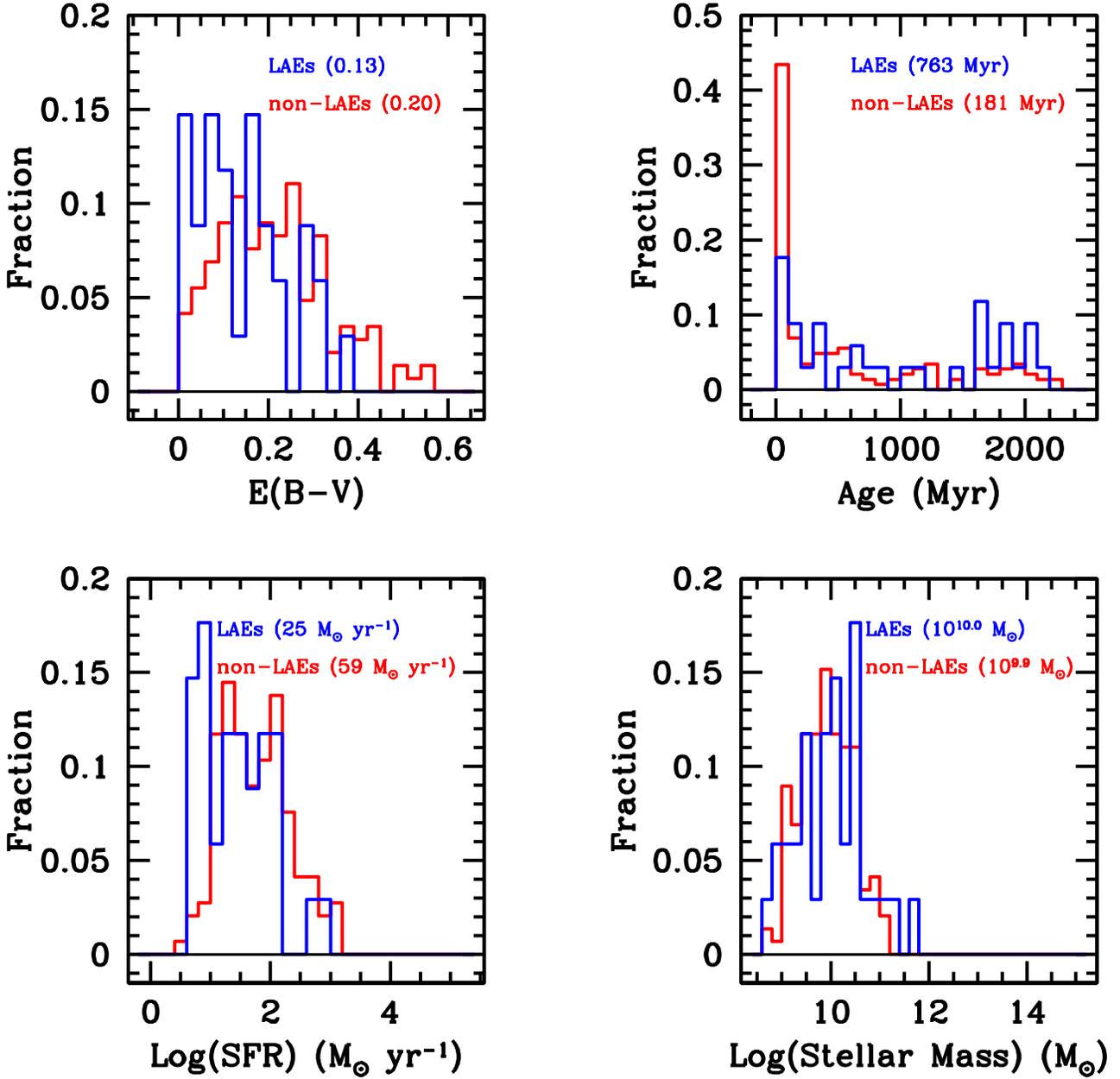}
\caption{Histograms of best-fit dust attenuations, ages, star formation rates, and stellar masses for the 34 LAEs (blue line) and 145 non-LAEs (red line), non-inclusive of objects with $K_s$ upper limits. The ordinate is the fractional percentage of the sample and the median value of each sample is indicated. The LAE sample is characterized by lower dust attenuation levels than the non-LAE sample; all objects with E(B-V) $\ge$ 0.40 are non-LAEs. The median age (763 Myr) of LAEs is significantly older than that of the non-LAEs (181 Myr). Furthermore, the LAEs lack the conspicuous overdensity of objects in the youngest ($t_{\star}$ $<$ 100 Myr) age bin that characterizes the non-LAE distribution. While the fraction of LAEs and non-LAEs with star formation rates between $\sim$ 80--120 $M_{\odot}$ yr$^{-1}$ is roughly the same, LAEs show a larger fraction of objects with small SFRs and non-LAEs exhibit a tail of high star formation rate objects. LAEs are marginally more massive than non-LAEs, on average, although the distributions are similar.}
\label{fig: LAE_nLAE_histo} 
\end{figure*}

We now turn to discussing our adoption of the \citet{bc2003} population synthesis models. While other models more fully take into account the Thermally-Pulsing Asymptotic Giant Branch (TP-AGB) stellar phase \citep[e.g.,][]{maraston2005}, a study of young star-forming ``BzK" galaxies at \z2 by \citet{daddi2007} found that the \citet{maraston2005} and \citet{bc2003} models yielded consistent M/L ratios; the LBGs in our sample are as least as dominated by the young stellar component as the \citet{daddi2007} BzK objects. Furthermore, a significant fraction ($\sim$ 40\%) of our sample does not have IRAC coverage, and therefore our SED fitting only includes photometry out to the rest-frame $V$ band (observed $K_s$ band) where the effects of TP-AGB stars on the derived parameters are less pronounced. Additionally, and perhaps most importantly, we are concerned with trends in the data (rather than specific stellar population values). It is these trends that are preserved regardless of model choice; \citet{lai2008} find that even though the \citet{maraston2005} models yield lower stellar masses and younger ages than \citet{bc2003} models for two samples of objects, the distinction between the two samples is significant irrespective of which population synthesis model is adopted. Finally, we note that our population synthesis results are supported by empirical color differences (\S \ref{sec: photometry}). In this sense, the choice of synthesis code is rendered less important -- there are clearly intrinsic differences between objects. 

In the next section, we divide the LBGs into two groups according to \lya strength and compare the stellar populations of these subgroups in order to more generally comment on the differences between objects with and without strong \lya emission. 

\subsection{The Distinct Properties of Strong \lya Emitters in the LBG Sample}

From the perspective of examining \lya emission, the simplest division of our data is according to \lya equivalent width. We turn here to comparing the properties of objects with strong ($W_{\rm Ly\alpha}$ $\ge$ 20 \AA) \lya emission and with weak \lya emission, or the line in absorption ($W_{\rm Ly\alpha}$ $<$ 20 \AA). Given our large sample size, these two groups each have sufficient membership to ensure robust statistics even when we restrict our analysis to only those objects with both stellar population modeling and no $K_s$ upper limits (34 and 145 objects, respectively). We call these two groups LAEs and non-LAEs, respectively, where we differentiate these objects from LAEs isolated with narrowband filters by explicitly referring to the latter as narrowband-selected objects. The motivation for choosing 20 \AA\ as the delineating equivalent width stems from the adoption of this value as a typical limit in narrowband LAE studies \citep[e.g.,][]{gawiser2007,nilsson2009,pentericci2009}. In order to investigate how stellar populations vary between LAEs and non-LAEs, we examined the age, E(B--V), stellar mass, and SFR distributions of both samples (Figure \ref{fig: LAE_nLAE_histo}). We discuss below the striking differences between the properties of UV continuum-bright LAEs and non-LAEs.

There is a remarkable dissimilarity between the age histograms of LAEs and non-LAEs: LAEs are generally fit with older best-fit ages than non-LAEs: LAEs have a median age more than four times that of the non-LAEs (763 Myr versus 181 Myr). LAEs furthermore lack the conspicuous overdensity of young objects ($t_{\star}$ $<$ 100 Myr) that characterizes the non-LAE sample. The disparity in ages between LAEs and non-LAEs is reflected in the K--S test probability of $\sim$ 2\% that the two age samples are drawn from the same underlying population. 

\begin{figure}[t]
\centering
\includegraphics[trim=0in 0in 0in 0in,clip,width=3.5in]{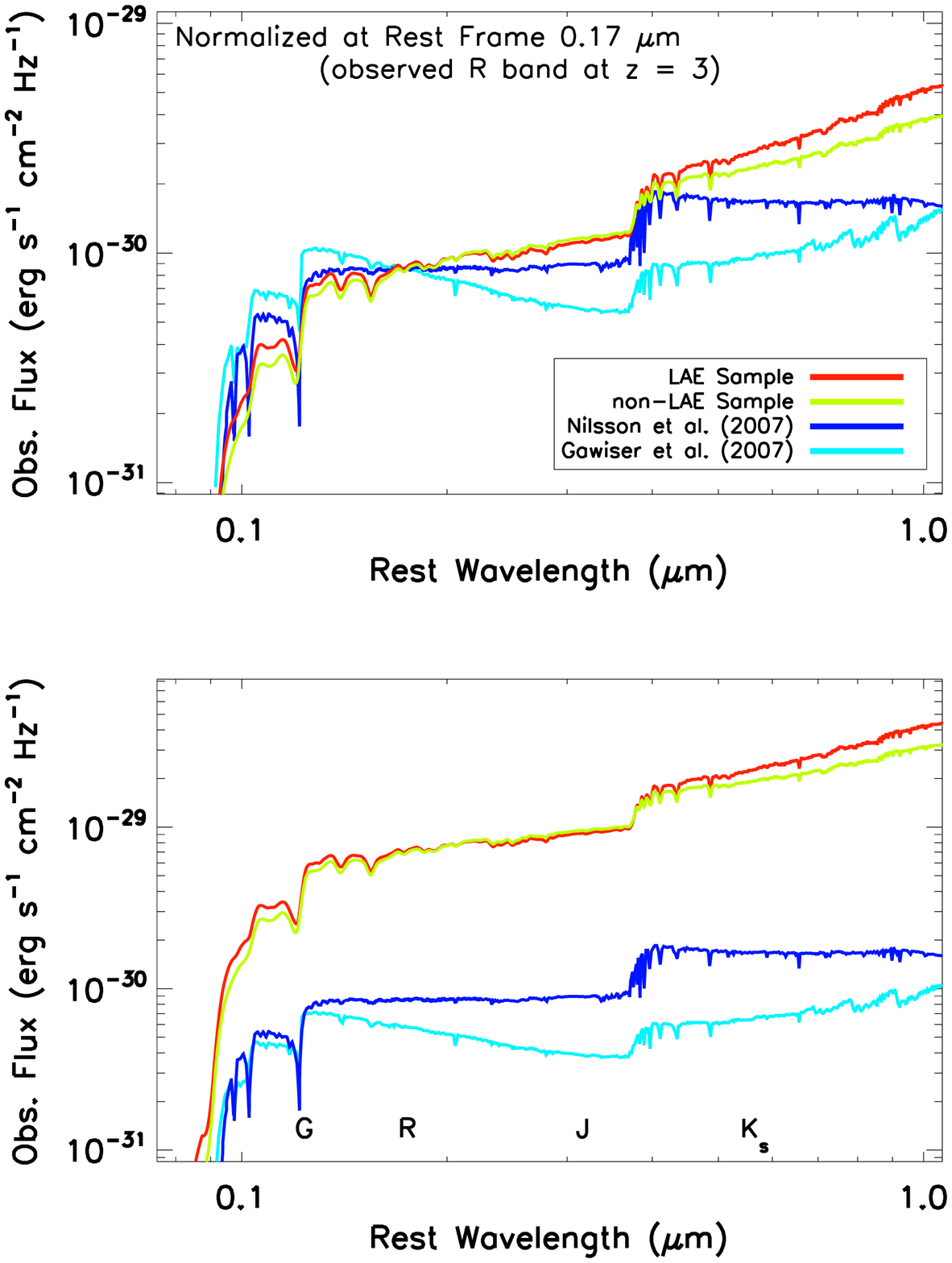}
\caption{\emph{Top panel:} Composite rest-frame UV to near-infrared SEDs of the LAE and non-LAE samples, assembled from best-fit SEDs from population synthesis modeling (omitting galaxies with $K_s$ upper limits). Data from \z3.1 narrowband-selected LAEs \citep{nilsson2007,gawiser2007} are overplotted and all spectra have been normalized to the \citet{nilsson2007} flux value at 0.17 $\mu$m (observed ${\cal R}$ band at \z3). Note that the LBG-selected objects have significantly redder $G$ -- ${\cal R}$ and ${\cal R}$ -- $K_s$ colors than the narrowband-selected LAEs, consistent with the older ages and higher dust attenuations derived for LBGs in the literature. \emph{Bottom panel:} Un-normalized spectra; note the significant flux discrepancy ($\sim$ 2.3 magnitudes) between the color-selected LBGs and the narrowband-selected LAEs.}
\label{fig: GawiserNilsson} 
\end{figure}

The distributions of dust attenuations for the LAE and non-LAE samples differ significantly, with a median E(B--V) of 0.13 for the LAEs and 0.20 for the non-LAEs. While both populations have roughly the same distribution of E(B--V) values below 0.30, the non-LAE distribution is characterized by a tail of high extinction values. No LAEs have E(B--V) $>$ 0.40, whereas 12/145 (8\%) of non-LAEs do. The K--S test probability that the LAE and non-LAE extinction distributions derive from the same parent population is $\sim$ 2\%.

The star formation rates of LAEs and non-LAEs are markedly different, as the LAEs have a median rate of 25 $M_{\odot}$ yr$^{-1}$ and the non-LAEs are characterized by a median rate of 59 $M_{\odot}$ yr$^{-1}$. The relative quiescence of strong Ly$\alpha$-emitters is in contrast to the high SFR tail observed in non-LAEs, where star formation rates in excess of 500 $M_{\odot}$ yr$^{-1}$ are recorded for 22 objects\footnote{15/22 of these objects, however, are younger than 10 Myr. While these objects have unphysically young ages \citep[given that the dynamical timescale of LBGs is $\ge$ 10 Myr;][]{giavalisco1996b,pettini1998,pettini2001}, re-modeling these galaxies with the constraint that $t_{\star}$ $\ge$ 10 Myr still results in best-fit star formation rates of several hundred $M_{\odot}$ yr$^{-1}$.}. There is a $\sim$ 1\% probability of these LAE and non-LAE star formation rate distributions are drawn from the same underlying population. 

Unlike age, E(B--V), and SFR, the stellar mass distributions of LAEs and non-LAEs are not strongly dissimilar. The median value of the LAEs (1.1 $\times$ 10$^{10}$ $M_{\odot}$) is only 40\% larger than that of the non-LAEs (7.9 $\times$ 10$^9$ $M_{\odot}$) and a K--S test predicts that the two distributions have a $\sim$ 27\% chance of being drawn from the same parent distribution. 

\begin{deluxetable*}{lllll}
\tablewidth{0pc}
\tablewidth{0pt}
\tablecaption{\textsc{Average Photometry}\tablenotemark{*}\label{table: phot}}
\tablehead{
  \multicolumn{1}{c}{Sample ($\langle{z} \rangle$)}
& \multicolumn{1}{c}{$\langle{{\cal R}}\rangle$ ($\sigma$)}
& \multicolumn{1}{c}{$\langle{G}\rangle$\tablenotemark{a} ($\sigma$)}
& \multicolumn{1}{c}{$\langle{G_{\rm IGM~corr.}}\rangle$ ($\sigma$)}
& \multicolumn{1}{c}{$\langle{K_s}\rangle$\tablenotemark{b}}
}
\startdata
LAEs (2.99) & 24.31 (0.02) & 24.85 (0.03) & 24.67 (0.03) & 21.24 \\
non-LAEs (2.97) & 24.34 (0.01) & 24.99 (0.01) & 24.81 (0.01) & 21.36 \\
\enddata
\tablenotetext{*}{Galaxies with $K_s$ upper limits were omitted from this analysis, as were objects lacking stellar population modeling.}
\tablenotetext{a}{$G$ magnitudes have been corrected to account for contributions from Ly$\alpha$.}
\tablenotetext{b}{Photometry calculated from best-fit SED, for objects lacking $K_s$ imaging.}
\end{deluxetable*}

We have shown that three of the best-fit stellar population parameters -- age, E(B--V), and SFR -- have markedly different distributions in the LAE and non-LAE samples. LAEs are typically older, less dusty, and less vigorously forming stars than non-LAEs. We tested the robustness of these results by constructing bootstrap samples from the confidence intervals of best-fit stellar parameters (\S \ref{sec: bootstrap}). The mean and the error on the mean ($\sigma$/$\sqrt n$, where $n$ is the sample size) were calculated for each best-fit stellar population parameter in the LAE and non-LAE samples. This process was repeated for 100 bootstrap samples. For E(B--V), age, and SFR, we found that the means of the LAE and non-LAE samples were consistently offset by at least their errors. In terms of stellar mass, the LAEs samples had consistently larger values; we note, however, that the masses of the LAE and non-LAE samples were comparable within their errors. These results, in agreement with the histograms discussed above, support the observed correlation that galaxies with strong \lya emission tend to be older, less dusty, and more quiescent than objects with weak \lya emission, or the line in absorption.

\subsection{Spectral Energy Distributions} \label{sec: photometry}

\begin{figure*}[t]
\centering
\includegraphics[trim=0in 0in 0in 0in,clip,width=5.5in]{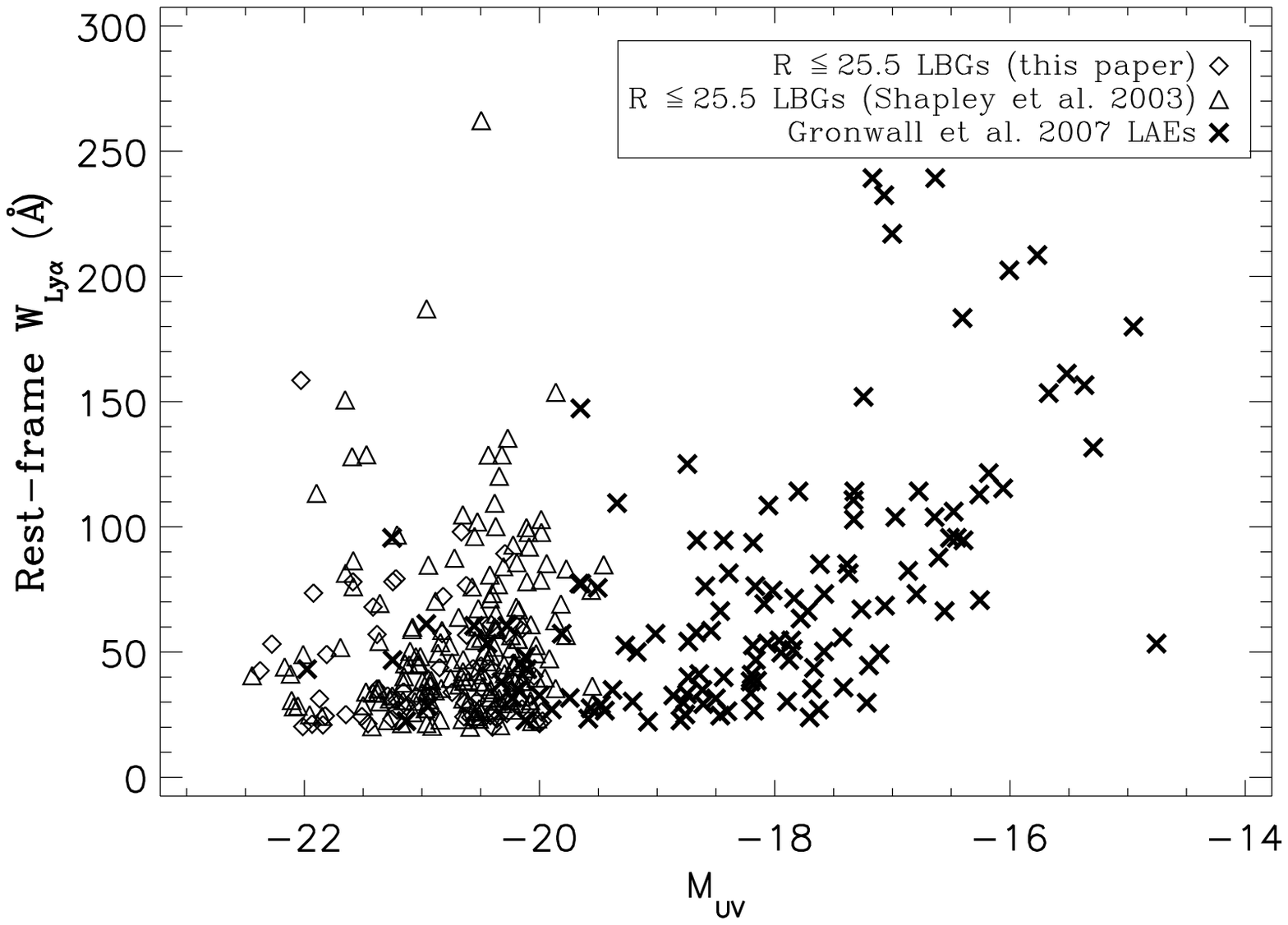}
\caption{Absolute UV magnitude versus rest-frame \ewlya\, for ${\cal R}$ $\le$ 25.5, \ewlya\ $\ge$ 20 \AA\ LBGs (diamonds and triangles) and the \citet{gronwall2007} sample of $z$ = 3.1 LAEs (crosses). Note that the LBG data include objects from \citet{shapley2003} and therefore span a larger range of \ewlya\ than that presented in this paper. A strong correlation between $M_{\rm UV}$ and \ewlya\ is observed in the \citet{gronwall2007} sample while weaker correlations are seen in the sample of \z3 LBGs. We attribute this discrepancy to the larger range of $M_{\rm UV}$ probed by LAEs. Object Q1009-C9 is both anomalously bright ($M_{\rm UV}$ = --22.0) and the strongest \lya emitter in our sample ($W_{\rm Ly\alpha}$ = 158.5 \AA), yet this galaxy does not exhibit any spectral signatures indicative of an AGN.}
\label{fig: MUV_EW_KSG} 
\end{figure*}

In order to place the results of the previous section on a more empirical footing, we transition here from focusing on stellar parameters and instead examine more broadly the relative colors of objects with and without strong \lya emission. We also contrast our data with SEDs of narrowband-selected LAEs from \citet{gawiser2007} and \citet{nilsson2007} with the aim of investigating the color differences, if any, between objects isolated with color cuts and narrowband filters. Our analysis is focused on the 179 objects in the LBG sample satisfying the criteria of population synthesis modeling and no $K_s$ upper limits. This empirical approach is complementary to the stellar population modeling discussed previously, and is furthermore a non-parametric probe of the differences between objects with and without strong \lya emission. Such a study is necessary given the well-known degeneracies among physical properties (e.g., dust extinction and age) derived from stellar population modeling \citep{shapley2003}. 

We find stark differences between the colors of LAEs and non-LAEs (Table \ref{table: phot}), where all $G$ magnitudes have been corrected for the contribution of \lya (\S \ref{sec:pops}). Strong Ly$\alpha$-emitters have bluer $G$ -- ${\cal R}$ colors than non-LAEs ($\langle G-{\cal R} \rangle_{\rm LAE}$ = 0.54; $\langle G-{\cal R} \rangle_{\rm non-LAE}$ = 0.65) and this trend is consistent when a correction for IGM absorption \citep{madau1995} is applied to the objects' $G$ magnitudes ($\langle G_{\rm IGM~corr.}-{\cal R} \rangle_{\rm LAE}$ = 0.36; $\langle G_{\rm IGM~corr.}-{\cal R} \rangle_{\rm non-LAE}$ = 0.47). Given that non-LAEs have younger best-fit stellar ages than LAEs (\S \ref{sec: bootstrap}), we expect that non-LAEs should have intrinsically bluer stellar continua than objects with strong \lya emission. The fact that we instead observe redder $G$ -- ${\cal R}$ colors in these objects is evidence that non-LAEs are more strongly attenuated by dust than LAEs. This result is consistent with the larger E(B--V) values derived for the non-LAEs than the LAEs (\S \ref{sec: bootstrap}). 

LAEs and non-LAEs also differ in their ${\cal R}$ -- $K_s$ colors, where $K_s$ magnitudes were inferred from best-fit SEDs for the 78 objects lacking near-infrared photometry. We find that LAEs are characterized by $\langle {\cal R}-{K_s} \rangle$ = 3.07, whereas non-LAEs are bluer by 0.1 magnitudes with $\langle {\cal R}-{K_s} \rangle$ = 2.98. Given that the ${\cal R}$ and $K_s$ passbands bracket the Balmer break for objects at \z3, a redder ${\cal R}$ -- $K_s$ color is indicative of a larger Balmer break. A more pronounced Balmer break, in turn, is correlated with an older stellar population \citep[e.g.,][]{bc2003}. Therefore, the result that LAEs have redder ${\cal R}$ -- $K_s$ colors than non-LAEs is empirical, non-parametric evidence that strong Ly$\alpha$-emitters are older on average. We have accordingly shown that the different colors of LAEs and non-LAEs are reflective of the trends discerned using stellar population modeling.}
 
Composite best-fit SEDs of the LAE and non-LAE samples were assembled by directly summing the individual best-fit SEDs over the rest frame range 912 \AA--1.0 $\mu$m and normalizing by the number of constituent spectra (34 and 145, respectively). In Figure \ref{fig: GawiserNilsson}, we compare the composite LAE and non-LAE spectra with best-fit average SEDs from two studies of narrowband-selected LAEs. Data from \citet{nilsson2007} and \citet{gawiser2007}, where the former include LAEs at \z3.1 in the GOODS South Field and the latter encompass LAEs at \z3.1 in the Extended Chandra Deep Field South, are shown both normalized and un-normalized with our LAE and non-LAE composite spectra. It is immediately apparent that the LBG-selected LAEs and non-LAEs are significantly brighter than the narrowband-selected LAEs ($\sim$ $\times$2.3 magnitudes in the ${\cal R}$ band). The LBG-selected objects also have redder $G$ -- ${\cal R}$ and redder ${\cal R}$ -- $K_s$ colors than both the \citet{nilsson2007} and \citet{gawiser2007} data. These differences in color are consistent with LBG-selected objects being both dustier and older than objects isolated with narrowband filters, as has been suggested by some authors \citep[e.g.,][]{gawiser2007,nilsson2009}. The relationship between LBGs and narrowband-selected LAEs is far from clear, though; some authors hypothesize a continuity between at least portions of the two populations based on similarities in stellar mass, color, and clustering \citep[e.g.,][]{adelberger2005a,gronwall2007,gawiser2007,lai2008,verhamme2008} whereas others maintain that the large luminosity and \ewlya\ discrepancies between LBGs and objects isolated with narrowband filters dictate separate evolutionary paths \citep[e.g.,][]{malhotra2002}. 

\section{Discussion} \label{sec: disc}

Our large data set of both spectroscopically-determined \lya equivalent widths and stellar population parameters constitutes a unique sample with which to investigate some of the current trends reported for  continuum- and narrowband-selected objects. We turn here to examining several of the outstanding questions pertaining to LBGs and narrowband-isolated LAEs, with the constant aim of physically motivating our work. We begin below with a discussion of the necessity of comparing objects drawn from the same parent luminosity distribution. 

\subsection{A Caveat: Differing Rest-Frame Luminosities} \label{sec: caveat}

By virtue of broadband selection techniques, LBGs have characteristically brighter optical continua than narrowband-selected LAEs (${\cal R}$ $\leq$ 25.5 versus $R$ $\sim$ 27). As it is critical to disentangle how luminosity and stellar populations are related before attributing galaxy differences purely to the same mechanisms that modulate \lya emission, it is of interest to investigate how, if at all, the derived properties of narrowband-selected LAEs are a natural result of preferentially isolating continuum-faint, line-bright objects. We compare our data with the work of several authors to explore the dependencies of \lya emission, stellar populations, and spatial clustering on broadband luminosity and we discuss the applications of such investigations to both LBGs and narrowband-selected LAEs. 

\emph{Luminosity and Equivalent Width:} Recent work has suggested a correlation between luminosity and \lya equivalent width, in the sense that few luminous LBGs have been observed with large equivalent widths \citep[e.g.,][]{shapley2003,erb2006a,ando2006,ouchi2008,vanzella2009,pentericci2009}. Several physical pictures have been suggested to explain this correlation, which has been noted in samples at redshifts $z$ $\sim$ 3--6. \citet{verhamme2008} and \citet{vanzella2009} hypothesized that different dust attenuations (from a luminosity-dependent chemical evolution history) could be responsible while \citet{ando2006} suggested that an enhanced presence of H\Rmnum{1} gas around luminous LBGs could explain the lack of large \lya equivalent widths in bright galaxies (consistent with the prediction of more massive galaxies residing in larger dark matter halos that are presumably richer in H\Rmnum{1} gas). Some authors, however, have conversely reported that the correlation between luminosity and \lya equivalent width is tenuous, at best \citep[e.g.,][]{steidel2000,verma2007,stanway2007,nilsson-prep}. Given the ongoing debate about such a correlation, we examined our extensive data set to test for a deficiency of luminous objects with large \lya equivalent widths. 

We restricted our analysis to the 64 objects with \ewlya\ $\ge$ 20 \AA\ in order to ensure completeness as a function of ${\cal R}$ magnitude, given that spectroscopic follow-up is biased toward objects with large, positive \lya equivalent widths. These objects have \ewlya\ $<$ 160 \AA\ and $M_{\rm UV}$ ranging from --20.0 to --22.4 (neglecting corrections for dust attenuation). This parameter space is comparable to that of \citet{ando2006}, where these authors suggest a threshold luminosity of $M_{\rm 1400}$ = --21.5 to --21.0 above which objects are deficient in large \ewlya\ values. We find that $M_{\rm UV}$ and \ewlya\ are not correlated in our sample, where the Spearman $\rho$ test statistic, $r_{\rm SR}$, is --0.104 and the probability of a null hypothesis, $P_{\rm SR}$, is 0.4076. Furthermore, we see no evidence for any such threshold luminosity in our data; when the sample is divided into bright ($M_{\rm UV}$ $<$ --21.0) and faint ($M_{\rm UV}$ $>$ --21.0) groups, the median \ewlya\ of each group are identical ($\sim$ 32 \AA).

These results are contradictory to those presented by \citet{shapley2003} who also examined a sample of \z3 LBGs. The \citet{shapley2003} sample, when limited to objects with \ewlya\ $\ge$ 20 \AA, consisted of objects with absolute UV luminosities between --19.5 and --22.4 (neglecting corrections for dust attenuation). In those data, there is a significant correlation between $M_{\rm UV}$ and \ewlya\ ($r_{\rm SR}$ = 0.274; $P_{\rm SR}$ = 0.0002). Dividing this sample into bright and faint subgroups, where the boundary luminosity is $M_{\rm UV}$ = --21.0, also yielded a pronounced difference in the median \ewlya\ values of each group: $M_{\rm UV}$(bright) = 35 \AA\ and  $M_{\rm UV}$(faint) = 44 \AA. Given that both our present sample and the \citet{shapley2003} sample are composed of \z3 LBGs, we hypothesize that one of reasons our data may not exhibit a correlation between $M_{\rm UV}$ and \ewlya\ is that our sample includes only a limited number of high equivalent width objects. Nine objects have \ewlya\ $\ge$ 70 \AA\ and just one has \ewlya\ $\ge$ 100 \AA\ in our data set whereas, in the \citet{shapley2003} sample, 45 objects have \ewlya\ $\ge$ 70 \AA\ and 20 have \ewlya\ $\ge$ 100 \AA. After testing that our systematic approach to calculating \lya equivalent width returned values consistent with those reported by \cite{shapley2003}, we combined our data with the 189 objects\footnote{We limit our analysis to the 179/189 galaxies satisfying the criterion \ewlya\ $\ge$ 20 \AA, measured according to our methodology described in \S \ref{sec: EW}; 10 objects classified as LAEs in the \citet{shapley2003} sample do not have equivalent widths above 20 \AA\ when measured using our systematic technique.} from \citet{shapley2003} with \ewlya\ $\ge$ 20 \AA. This combined sample of 243 objects, with equivalent widths measured in a uniform manner, was then tested for a correlation between $M_{\rm UV}$ and \ewlya. We found only a moderate correlation ($r_{\rm SR}$ = 0.149; $P_{\rm SR}$ = 0.0202), where we hypothesize that the trend between $M_{\rm UV}$ and \ewlya\ is simply not strong enough to be unequivocally detected in all samples. Furthermore, the correlation between $M_{\rm UV}$ and \ewlya\ in these data may be masked by the relatively small range of $M_{\rm UV}$ probed by LBGs. Narrowband-selected LAEs exhibit a wider range of absolute luminosities (23.0 $\lesssim$ ${R}$ $\lesssim$ 27.0) than LBGs do (23.0 $\lesssim$ ${\cal R}$ $\lesssim$ 25.5); any trend between $M_{\rm UV}$ and \ewlya\ will consequently be more pronounced in a sample of LAEs. In order to test this statement, we analyzed a sample of LAEs from \citet{gronwall2007}. We divided the data into two samples, where the first subsample included only objects brighter than the LBG spectroscopic limit and the second subsample was inclusive of all the data. The first subsample showed no correlation between $M_{\rm UV}$ and \ewlya\ ($r_{\rm SR}$ = --0.015; $P_{\rm SR}$ = 0.9185) whereas the second subsample exhibited a strong correlation ($r_{\rm SR}$ = 0.585; $P_{\rm SR}$ = 0.0000). These results support our hypothesis that the trend between $M_{\rm UV}$ and \ewlya\ at \z3 is apparent only when considering data encompassing a large range in absolute luminosity. In Figure \ref{fig: MUV_EW_KSG}, we present a plot of $M_{\rm UV}$ versus \ewlya\ for ${\cal R}$ $\le$ 25.5, \ewlya\ $\ge$ 20 \AA\ LBGs and the \citet{gronwall2007} sample of narrowband-selected LAEs. Even if one assumes incompleteness in the \citet{gronwall2007} sample at the faint end, the lack of large equivalent widths among bright objects is striking \citep[but see][]{nilsson-prep}. \citet{ando2006} report a deficiency of bright objects with large equivalent widths in their sample of \emph{z} = 5 -- 6 objects; that fact that we observe several such objects here while those authors do not can be attributed to the small sample size of \citet{ando2006}. 

\emph{Luminosity and Stellar Populations:} Understanding how luminosity and stellar populations are correlated is critical to commenting on the nature of LBGs and LAEs, given that the latter are commonly an order of magnitude fainter in optical broadband filters. \citet{lai2008} examined a sample of 70 \z3.1 LAEs, $\sim$ 30\% of which were detected in the $Spitzer$ IRAC 3.6 $\mu$m band. These authors found that the IRAC-detected LAEs were significantly older and more massive ($\langle t_{\star} \rangle$ $\sim$ 1.6 Gyr, $\langle M \rangle$ $\sim$ 9 $\times$ 10$^{9}$ $M_{\odot}$) than the IRAC-undetected sample ($\langle t_{\star} \rangle$ $\sim$ 200 Myr, $\langle M \rangle$ $\sim$ 3 $\times$ 10$^8$ $M_{\odot}$). In addition to having redder colors, the IRAC-detected sample was also approximately three times brighter in the $R$ band. This correlation between luminosity and redness, age, and mass prompted \citet{lai2008} to suggest that the IRAC-detected LAEs may be a low-mass extension of the LBG population. The relationship between luminosity and average dust attenuation was examined by \citet{reddy2008}, where these authors studied a sample of UV-selected objects at 1.9 $\le$ $z$ $<$ 3.4 and found no correlation between $M_{\rm UV}$ and E(B--V) over the absolute magnitude range 22.0 $\le$ ${\cal R}$ $\le$ 25.5. While bright objects do not appear to exhibit a correlation between dust attenuation and luminosity, \citet{reddy2008} postulate that objects fainter than the LBG spectroscopic limit of ${\cal R}$ = 25.5 may have lower average dust attenuations than brighter objects. This has been confirmed by \citet{bouwens-prep}, who examined a sample of objects at $z$ $\sim$ 2--6 over a larger dynamic range (0.1$L^*_{z \rm=3}$ to 2$L^*_{z \rm=3}$) than had been previously studied. These authors reported a correlation between UV luminosity and dust attenuation such that fainter objects are bluer. Given the correlation we find between E(B--V) and \ewlya, the evolution of dust attenuation with UV luminosity may have implications for the fraction of LAEs as a function of UV luminosity \citep{reddy2009}. These results have highlighted that examining objects over only a limited range of absolute magnitudes may mask an underlying correlation. 

\emph{Luminosity and Clustering:} Several authors have noted the different clustering properties of LBGs and LAEs, where the former exhibit a mean halo occupation of $\sim$ 100\% \citep[e.g.,][]{conroy2008} in dark matter halos with minimum masses of $\sim$ 10$^{11.3}$ $h^{-1}$ $M_{\odot}$ \citep[e.g.,][]{adelberger2005a,conroy2008} and the latter cluster more weakly and appear to occupy only 1/100 -- 1/10 of similarly-clustered dark matter halos with lower limit masses of $\sim$ 10$^{10.6}$ $h^{-1}$ $M_{\odot}$ \citep[e.g.,][]{gawiser2007}. Given the direct dependence of clustering strength on UV luminosity \citep[e.g.,][]{ouchi2004,adelberger2005a,lee2006b}, the spatial differences between
LBGs and LAEs may simply reflect the discrepancy in typical UV luminosity between the two samples.  It is therefore of interest to investigate how the clustering strength of LAEs correlates with \lya luminosity or \lya equivalent width, given that these quantities govern the selection of LAEs. Discerning the relative clustering properties of objects with and without strong \lya emission {\it within the same UV luminosity range} is critical to understanding the correlation between \lya and galactic properties. 

These results -- that objects more luminous in the rest-frame UV may \emph{1)} be deficient in large \lya equivalent widths and \emph{2)} be generally redder, older, and more massive relative to less luminous objects -- may be the driving factors responsible for (faint, high equivalent width) narrowband-selected LAEs being typically younger, bluer, and less massive than (bright, lower equivalent width) LBGs. 

As the LBG spectroscopic limit of ${\cal R}$ $\le$ 25.5 necessarily limits our sample to a smaller absolute magnitude range than than probed by narrowband-selected LAEs (23.0 $\lesssim$ ${R}$ $\lesssim$ 27.0), the luminosity-dependent trends reported above for \lya equivalent width, stellar populations, and clustering \citep[e.g.,][]{ando2006,lai2008,adelberger2005a} may only become apparent in narrowband-selected samples where the absolute luminosity range is significantly larger than it is in samples of LBGs. We remind the reader that while our conclusions are applicable to both LBGs and bright (${\cal R}$ $\le$ 25.5) narrowband-selected LAEs, we are unable to make inferences about the population of faint LAEs. We refer the reader to \citet{lai2008} for a discussion of these objects. 

\subsection{LBG and LAE Equivalent Width Distributions}

The equivalent width of the \lya feature is a common benchmark that can be used to make comparisons between studies of LBGs and LAEs. Observing programs of flux-selected Ly$\alpha$-emitters typically employ a narrowband ($\sim$ 50--100 \AA) filter centered on the redshifted \lya line, paired with a broadband filter used to characterize the local continuum. Samples are defined according to a minimum \lya equivalent width -- typically $\sim$ 20 \AA\ in the rest frame -- where \ewlya\ is calculated photometrically by ratioing the flux in the narrowband filter to the flux density in the broadband filter. LBGs, on the other hand, are selected according to broadband flux and color cuts, with no implicit requirement on \ewlya. Equivalent widths of LBGs are most commonly calculated spectroscopically, as these objects are characterized by relatively bright continua (${\cal R}$ $\leq$ 25.5). Given these different selection techniques and methods of calculating \ewlya, it is of interest to compare the equivalent width distributions of LBG-selected LAEs and narrowband-selected LAEs. A difference in the \ewlya\ distributions of these two populations would be indicative of an underlying dissimilarity between LBG-selected LAEs and narrowband-selected LAEs, even though both classes of objects can be broadly classified as strong Ly$\alpha$-emitters\footnote{\citet{reddy2008} found that the standard LBG color selection criteria (Equation \ref{color-cuts}) do little to bias the intrinsic distribution of \ewlya\ for \z3 LBGs, given the low probability of an object having a strong enough \lya feature to perturb its colors out of the LBG selection window. We therefore attribute any differences in the \ewlya\ distributions of LBGs and narrowband-selected LAEs to intrinsic dissimilarities between the two populations.}.

For this analysis, we combined our present sample of 64 LBG-selected LAEs with the 179 objects from \citet{shapley2003} satisfying the criterion of \ewlya\ $\ge$ 20 (measured using our methodology described in \S \ref{sec: EW}). Integrating these two data sets yields a large sample of \z3 objects for which equivalent widths were measured in an identical manner. We find that these data are characterized by a median equivalent width of 42 \AA\ and mean of 57 \AA. In contrast, a sample of 160 LAEs at $z$ = 3.1 in the Extended Chandra Deep Field South observed by \citet{gronwall2007} had a median of $\sim$ 60 \AA\ and a mean of $\sim$ 80 \AA. We used the K--S test to quantify the likelihood that the LBG-selected LAEs and the narrowband-selected LAEs derived from the same parent equivalent width distribution. When the entire sample of objects from \citet{gronwall2007} was considered, we found a probability of 3.5 $\times$ 10$^{-4}$ that the objects shared a common equivalent width distribution. Alternatively, when we limited the \citet{gronwall2007} data to include only objects satisfying the LBG spectroscopic limit of ${\cal R}$ $\le$ 25.5, we obtained a probability of $\sim$ 0.95 that the equivalent widths of the LBG- and narrowband-selected LAE samples could be described as originating from the same population. This result, consistent with \citet{verhamme2008}, is indicative of the similarity in equivalent width of objects spanning comparable UV luminosities. Conversely, the probability that the equivalent widths of our sample of LBG-selected LAEs and the faint (${R}$ $\ge$ 25.5) \citet{gronwall2007} LAEs derive from the same parent population is 8 $\times$ 10$^{-8}$. Similarly, when the \citet{gronwall2007} data are divided using the boundary $R$ = 25.5, a K--S test yields a probability of 7.9 $\times$ 10$^{-6}$ that the two equivalent width populations derive from the same parent distribution.

In other words, while the equivalent width distributions of LBG- and narrowband-selected LAEs are quite disparate when considered at face value, it appears that the bulk of this disparity can be attributed to the  different luminosities probed by the respective samples. When narrowband-selected LAEs are restricted to objects more luminous than $R$ = 25.5, a K--S test reveals a high probability that the equivalent width distributions of strong Ly$\alpha$-emitting LBGs and bright LAEs derive from the same parent population. This similarity in equivalent width, when a standard continuum magnitude range is adopted, is consistent with the picture that bright LAEs represent the same population as Ly$\alpha$-emitting LBGs \citep[e.g.,][]{verhamme2008}. 

\subsection{Escape of \lya Photons}

As \lya is susceptible to a variety of complex radiation transport effects including resonant scattering \citep[e.g.,][]{neufeld1990} and the velocity field of the ISM \citep[e.g.,][]{verhamme2008}, the escape of \lya emission is a probe of the physical conditions of galaxies' interstellar media. In the simplest case of a single-phase ISM in which gas and dust are well mixed, \lya photons experience more attenuation than continuum photons due to the longer dust absorption path lengths expected for resonantly scattered radiation. To explain the large \lya equivalent widths observed in some apparently dusty LAEs, the presence of a multi-phase ISM -- in which dust is segregated in neutral clouds -- has been proposed \citep{neufeld1991,hansen2006,finkelstein2009}. In such a geometry, \lya photons are scattered at the surface of dusty clumps while continuum photons penetrate the clumps and are preferentially scattered or absorbed. Galaxies in which the dust is segregated in neutral clouds are consequently predicted to exhibit larger \lya equivalent widths than would be expected given their stellar populations. 

\begin{figure}[t]
\centering
\includegraphics[trim=0in 0in 0in 0in,clip,width=3.7in]{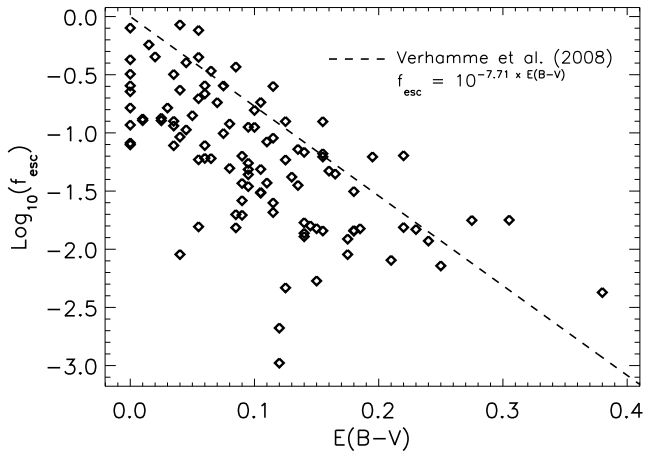}
\caption{\lya escape fraction as a function of dust extinction, for objects with best-fit stellar ages $t_{\star}$ $>$ 100 Myr. A significant correlation is observed, in the sense that objects with lower dust attenuations have larger escape fractions. The best-fit line from \citet{verhamme2008} is overplotted as a dashed line.}
\label{fig: f_esc} 
\end{figure}

We considered the likelihood of the above dust geometry in our sample of LBGs. Our methodology relied on comparing observed and predicted \lya luminosities, where the former, $L_{\rm Ly \alpha(flux)}$, was calculated simply from the \lya line flux and the latter, $L_{\rm Ly \alpha(SFR)}$, was derived from the conversion between SFR (from population synthesis modeling) and \lya luminosity, assuming the conventions (Case B recombination; $T_{\rm e}$ = 10,000 K) of \citet{kennicutt1998} and \citet{brocklehurst1971}:

\begin{equation} \label{eqn: Lya} L_{\rm Ly \alpha(SFR)} = \frac{\rm SFR ~[\it M_{\odot} \rm ~yr^{-1}]}{9.1 \times 10^{-43}} \end{equation}

\noindent where SFR was calculated assuming a \citet{salpeter1955} IMF. Two similar parameters incorporating observed and predicted \lya luminosities were of interest: the escape fraction, $f_{esc}$, and the relative escape fraction, $f_{esc,rel}$. The escape fraction is the ratio of observed to predicted \lya luminosities, whereas the relative escape fraction is this same ratio with an extra term corresponding to a dust correction at the rest wavelength of Ly$\alpha$. $f_{esc,rel}$ is a probe of the degree to which \lya and continuum photons experience the same level of dust attenuation. In the case that $f_{esc,rel}$ = 1, \lya and continuum photons are attenuated by the same E(B--V). On the other hand, a relative escape fraction less than unity is indicative of a dust geometry in which \lya photons are attenuated more than continuum photons. Conversely, a galaxy in which dust is segregated in neutral clouds and where the \lya flux is consequently enhanced relative to the continuum flux would have $f_{esc,rel}$ $>$ 1. We present the equations for  $f_{esc}$ and  $f_{esc,rel}$ below, where $k'$(1216) parameterizes the \citet{calzetti2000} starburst attenuation law at 1216 \AA:

\begin{equation} f_{\it esc} = \frac{ L_{\rm Ly\alpha(flux)}}{ L_{\rm Ly\alpha(SFR)}},\end{equation}

\begin{equation} f_{\it esc,rel} = f_{\it esc} \times 10^{\rm 0.4E(B-V)\it k'(\rm 1216)}. \end{equation}

When calculating $f_{esc}$ and $f_{esc,rel}$, we limited our analysis to only those objects with \ewlya\ $>$ 0 \AA\ due to the unphysical nature of inferring a \lya luminosity from a negative \lya flux. We also required objects to have best-fit stellar ages older than 100 Myr in order to ensure the validity of Equation \ref{eqn: Lya}, which assumes continuous star formation for at least that duration \citep{kennicutt1998}. These criteria isolated a sample of 105 objects. We note that the questions raised about the nature of the dust extinction law for young objects (\S \ref{sec: bootstrap}) are not of concern given this older sample. To correct for slit losses, we normalized the $G$ magnitudes determined from photometry and spectroscopy; the latter were calculated by multiplying the $G$ filter transmission curve (4780/1100 \AA) over the optical spectra. The \lya line fluxes were adjusted accordingly, where the median flux correction was an increase by a factor of $\sim$ 1.7 (0.6 magnitudes). This procedure relies on the assumption that \lya emission has the same spatial extent as broadband 940--1470 \AA\ emission. While some recent work \citep[e.g.,][]{steidel2000,matsuda2004,hayashino2004} has suggested that \lya and UV continuum emission may be spatially decoupled with \lya more extended than continuum emission, in the absence of simultaneous high-resolution \lya and UV continuum imaging for our entire sample we assume for simplicity that the line and continuum emission are coincident. 

The calculated values of $f_{esc}$ vary from 0.00--0.85. In Figure \ref{fig: f_esc}, we plot E(B--V) versus $f_{esc}$; we find a strong correlation in the sense that objects with large dust attenuations tend to have small \lya escape fractions ($r_{\rm SR}$= --0.712, $P_{\rm SR}$ =  0.0000). We note that this trend is consistent with the existence of LBGs in our sample exhibiting both \lya in absorption and red UV continua, where these particular objects were not included in this analysis due to their negative \lya fluxes. Other authors have derived similar results, although with significantly smaller sample sizes \citep[e.g.,][]{verhamme2008,atek2009}. The scatter present in our larger data set -- some of which may be due to systematics in estimating $f_{esc}$ -- is a useful probe of the nature of \lya radiative transfer in LBGs. We further discuss a physical picture of LBGs consistent with these $f_{esc}$ results in \S \ref{sec: sum}.

\begin{figure}[t]
\centering
\includegraphics[trim=0in 0in 0in 0in,clip,width=3in]{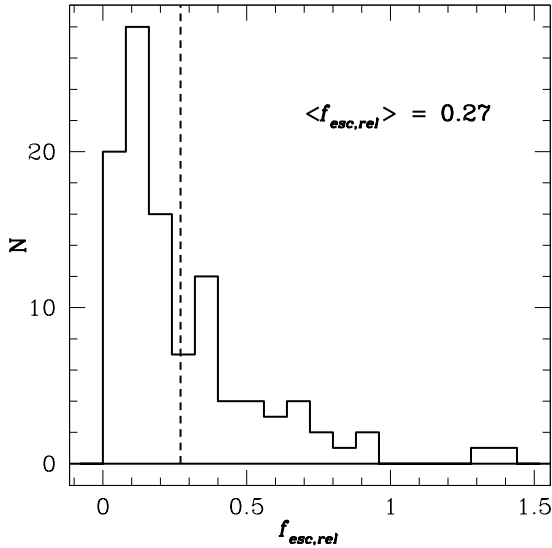}
\caption{Histogram of $f_{esc,rel}$ values, for objects with $t_{\star}$ $>$ 100 Myr. Note that the majority of objects (103/105) have $f_{esc,rel}$ $<$ 1, where the sample as a whole is characterized by $\langle f_{esc,rel} \rangle$ = 0.27. These low relative escape fractions are indicative of \lya photons experiencing more attenuation than non-resonance continuum photons. These results are not consistent with the hypothesis of a clumpy interstellar medium, as has been suggested to exist in some LAEs \citep[e.g.,][]{finkelstein2009}.}
\label{fig: f_rel_esc} 
\end{figure}

In order to investigate the nature of the gas and dust distributions for the objects in our sample, we derived the relative escape fraction of the 105 LBGs with \ewlya\ $>$ 0 \AA\ and $t_{\star}$ $>$ 100 Myr. We found that 103 objects had $f_{esc,rel}$ $<$ 1 and the sample as a whole was characterized by $\langle f_{esc}(\rm rel) \rangle$ = 0.27 (Figure \ref{fig: f_rel_esc}). For comparison, \citet{gronwall2007} and \citet{nilsson2009} report typical $f_{esc,rel}$ values of $\sim$ 0.30 and $\sim$ 0.60 in LAE samples at \z3.1 and \z2.25, respectively.  Even if we conservatively assume \lya slit losses for our sample approaching a factor of two \citep{hayashino2004}, these low relative escape fractions, well below unity, do not support the hypothesis in which dust is primarily segregated to neutral clouds in Ly$\alpha$-emitting LBGs. We note that the absorption of \lya photons by the IGM is not a plausible explanation for these low relative escape fractions; at \z3, \lya line fluxes are predicted to experience an IGM-induced decrement of only $\sim$ 20\% \citep{madau1995,shapley2006}. Furthermore, the observed symmetry of the \lya lines in our sample (Figures \ref{fig: redshift} and \ref{fig: ages}) are indicative of minimal IGM absorption. Additionally, if one adopts a sub-solar metallicity (in contrast to $Z_{\odot}$ used in this paper), the calculated values of $f_{esc,rel}$ will be \emph{lower} than those presented here due to the fact that the ratio of the ionizing photon luminosity to the observed UV luminosity increases with decreasing metallicity \citep[e.g.,][]{leitherer1999}. This result of relative escape fractions below unity indicates that \lya appears to experience more attenuation than continuum photons, consistent with the physical picture of a homogeneous distribution of gas and dust in which resonantly scattered \lya photons have longer dust absorption path lengths than continuum photons. Such a simple picture is likely not an accurate representation of the ISM of \z3 LBGs; indeed, the ISM of our own Milky Way is known to be highly inhomogeneous \citep{dickey1989}. However, our results suggest a scenario in which dust and gas are well mixed among the different phases of an inhomogeneous ISM. 

\section{Summary and Conclusions} \label{sec: sum}

We have analyzed a sample of 321 optically-selected \z3 Lyman break galaxies in order to investigate the relationship between stellar populations and \lya emission. The equivalent width of the \lya feature was robustly estimated from rest-frame UV spectroscopy and broadband $G{\cal R}JK_s$ + $Spitzer$ IRAC photometry was used to conduct stellar population modeling to derive the key properties of age, extinction, star formation rate, and stellar mass for all objects with photometric coverage in at least one near- or mid-infrared passband ($\sim$ 80\% of the sample). The limited luminosity range of LBGs enabled a controlled investigation of the nature of \lya emission \emph{within a sample of objects spanning similar luminosities.} Given the luminosity-dependent trends in \lya equivalent width, stellar populations, and clustering \citep[e.g.,][]{gronwall2007,lai2008,adelberger2005a}, this controlled study represents an improvement over previous work. We used a variety of statistical tests to analyze the correlation between \lya emission and stellar populations in LBGs from the standpoint of comparing spectroscopic observations, photometric data, and best-fit stellar population parameters. The relative \lya escape fraction was used to probe the relative distributions of gas and dust in the objects' interstellar media. Below, we summarize our results.

\begin{itemize}
\item \lya equivalent width is correlated with age, SFR, and E(B--V), respectively, in the sense that galaxies with strong \lya emission are older, more quiescent, and less dusty than their counterparts with weak or absent \lya emission. Taking into account the uncertainties on the best-fit population parameters, the probability of a null hypothesis (i.e., no correlation) was excluded at the 3--5$\sigma$ level for these relationships. We found that stellar mass was not significantly correlated with \ewlya\ -- the null hypothesis could be excluded at only the $\sim$ 1$\sigma$ level. These results were consistently derived from several different analyses of the data, including correlation tests, composite spectra, and the binary division of the sample into LAEs ($W_{\rm Ly \alpha}$ $\ge$ 20 \AA) and non-LAEs ($W_{\rm Ly \alpha}$ $<$ 20 \AA). 

\item Analysis of the relative escape fraction of Ly$\alpha$ is consistent with \lya photons experiencing more attenuation than non-resonance continuum photons. We also find that the \lya escape fraction is strongly correlated with E(B--V), where galaxies with more dust attenuation also have lower escape fractions. 
\end{itemize}

The observed correlations between \lya emission and stellar populations are consistent with the physical picture proposed by \citet{shapley2001}, in which young, dusty LBGs experience vigorous outflows from supernovae and massive star winds. \citet{shapley2003} reported evidence that both gas and dust are entrained in the outflows, evidence that these ``superwinds" could tenably decrease a galaxy's dust and gas covering fraction over several tens of Myr. While the more mature LBGs may have as much or more overall dust content than younger galaxies, the fact that these older objects ($t_{\star}$ $\gtrsim$ 100 Myr) are also more likely to exhibit \lya emission is an important point: this trend demonstrates that it is the dust covering fraction -- not the total amount of dust -- that modulates \lya emission. Independent evidence for a lower covering fraction of dusty gas in older objects is provided by a trend toward weaker low-ionization interstellar absorption lines with increasing age in the composite spectra presented in \S \ref{sec: bootstrap}.

While this physical picture has been put forth previously, our current results are more strongly supported due to the wide range of statistical tests and analysis methods that we employed. We compared our larger sample of data in different forms -- spectroscopic, photometric, and best-fit parameters -- using a battery of statistical methods including correlation and K--S tests. \lya emission strength was furthermore treated as both an independent and dependent variable. These analyses represent a systematic, powerful approach to elucidating trends. 

Results of both the escape fraction and the relative escape fraction can be used to further constrain the physical picture of LBGs and bright narrowband-selected LAEs. While the observed inverse relationship between E(B--V) and $f_{esc}$ is consistent with earlier results \citep[e.g.,][]{verhamme2008,atek2009}, the scatter present in our larger sample highlights the myriad factors modulating the \lya escape fraction (e.g., galaxy kinematics, dust, and outflow geometry). Further analysis of this scatter will be instrumental in constraining how the \lya escape fraction varies. The observed relative escape fraction of \lya emission -- where the majority of LBGs have $f_{esc,rel}$ below unity -- is indicative of a physical picture in which LBGs contain gas and dust that is well mixed. In this geometry, \lya photons experience more attenuation than continuum photons do because of the former's increased path lengths from resonant scattering. This is in contrast to the scenario proposed by \citet{finkelstein2009}, in which dust is segregated in neutral gas clumps surrounded by an ionized, dust-free medium. We caution, however, that the relative escape fraction of \lya is likely modified by a variety of factors including galaxy kinematics, dust, and outflow geometry. As such, additional observations are necessary in order to more fully understand the distribution (and evolution) of gas and dust in LBGs. 

While we hope that this work has illuminated the relationship between \lya emission and stellar populations in \z3 LBGs, future systematic studies at other redshifts -- ideally over larger dynamic ranges in UV luminosity -- are necessary in order to address the discrepancies between trends reported here and those at \z2 or \z4 \citep[e.g.,][]{erb2006a,pentericci2007}. The nature of complex \lya emission morphologies -- such as the double-peaked objects that comprise $\sim$ 4\% of the sample discussed here -- has also yet to be explored in light of understanding if these objects' stellar populations and gas distributions differ from those of more typical Ly$\alpha$-emitters or absorbers. Similarly, an analysis of the geometry of dust and gas in the interstellar media of \z3 LBGs is vital to understanding what kind of dust attenuation law -- the \citet{calzetti2000} relation, a SMC-like law, or another relation completely -- most accurately describes these objects. Furthermore, a basic property critical to placing galaxies in an evolutionary context -- metallicity -- remains poorly constrained for the \z3 LBG population \citep[but see][]{maiolino2008}. While metallicity has been investigated in the rare cases of gravitationally lensed objects \citep[e.g.,][]{teplitz2000,pettini2002,hainline2009,quider2009}, a systematic study of elemental abundances in a larger sample of \z3 LBGs would be illuminating \citep[akin to the \z2 sample discussed in][]{erb2006a}. Given anticipated advances in both telescopes and instrumentation, we hope that these questions are illustrative of many others that will help to guide future LBG research.

\begin{acknowledgements}
We thank Kim Nilsson and Eric Gawiser for kindly providing their data for comparison. A.E.S. acknowledges support from the David and Lucile Packard Foundation and the Alfred P. Sloan Foundation. D.K.E. is supported by a Spitzer Fellowship through a NASA grant
administrated by the Spitzer Science Center. C.C.S. has been supported by grants AST 03-07263 and AST 06-06912 from the National Science Foundation and by the David and Lucile Packard
Foundation. Support for N.A.R. was provided by NASA through Hubble Fellowship grant
HST-HF-01223.01 awarded by the Space Telescope Science Institute, which is
operated by the Association of Universities for Research in Astronomy,
Inc., for NASA, under contract NAS 5-26555. We also wish to recognize and acknowledge the very significant cultural role and reverence that the summit of Mauna Kea has always had within the indigenous Hawaiian community.  We are most fortunate to have the opportunity to conduct observations from this mountain. This work is based in part on observations made with the \emph{Spitzer Space Telescope}, which is operated by the Jet Propulsion Laboratory, California Institute of Technology under a contract with NASA. 
\end{acknowledgements}

\bibliography{lbgrefs}

\clearpage

\pagebreak

\end{document}